\documentclass[amsmath,eqsecnum,preprint,pre,aps,showpacs]{revtex4}
\usepackage{amsmath}
\usepackage{graphicx}
\usepackage{color}
\usepackage{dcolumn}
\usepackage{bm}
\usepackage{float}

\begin{document}

\draft
\emph{}
\title{Non-equilibrium  Theory of Arrested Spinodal Decomposition}
\author{Jos\'e Manuel Olais-Govea, Leticia L\'opez-Flores and
Magdaleno Medina-Noyola}

\address{Instituto de F\'{\i}sica {\sl ``Manuel Sandoval Vallarta"},
Universidad Aut\'{o}noma de San Luis Potos\'{\i}, \'{A}lvaro
Obreg\'{o}n 64, 78000 San Luis Potos\'{\i}, SLP, M\'{e}xico}

\date{\today}

\begin{abstract}
The Non-equilibrium Self-consistent  Generalized Langevin Equation
theory of irreversible relaxation [Phys. Rev. E (2010)
\textbf{82}, 061503; ibid. 061504] is applied to the description
of the non-equilibrium processes involved in the spinodal
decomposition of suddenly and deeply quenched  simple liquids. For
model liquids with hard-sphere plus attractive (Yukawa or square
well) pair potential, the theory predicts that the spinodal curve,
besides being the threshold of the thermodynamic stability of
homogeneous states, is also the borderline between the regions of
ergodic and non-ergodic homogeneous states. It also predicts that
the high-density liquid-glass transition line, whose
high-temperature limit corresponds to the well-known hard-sphere
glass transition,  at lower
temperature intersects the spinodal curve and continues inside the spinodal
region as a glass-glass transition line. Within the region bounded
from below by this low-temperature glass-glass transition and from
above by the spinodal dynamic arrest line we can recognize two
distinct domains with qualitatively different temperature
dependence \textcolor{black}{of various physical properties. We interpret these two domains as corresponding} to full
gas-liquid phase separation conditions and to the formation of physical gels by
arrested spinodal decomposition. \textcolor{black}{The resulting theoretical scenario is consistent with the corresponding experimental observations in a specific colloidal model system.}
\end{abstract}


\maketitle

\section{Introduction.}\label{sectionI}

Quenching a liquid from supercritical temperatures into the
liquid/gas coexistence region leads to the separation of the
system into the equilibrium coexisting phases
\cite{callen,mcquarrie}. This phase separation process is a
paradigmatic illustration of non-equilibrium irreversible
processes, whose description must be based on the general
fundamental principles of molecular non-equilibrium statistical
thermodynamics \cite{keizer,casasvazquez0}. When the final density
and temperature of the quench correspond to a state point $(n,T)$
inside the spinodal region, the separation process starts with the
amplification of spatial density fluctuations, eventually leading
to the two-phase state in thermodynamic equilibrium
\cite{cahnhilliard,cook,furukawa,langer,dhont,goryachev}. However,
under some conditions, involving mostly colloidal liquids
\cite{luetalnature,sanz,cardinaux,gibaud,foffi,Gao}, this process of phase
separation is interrupted when the colloidal particles cluster in
a percolating network of the denser phase, forming an amorphous
sponge-like non-equilibrium bicontinuous structure,  typical of
physical gels \cite{zaccarellireviewgels}.

In understanding these experimental observations, computer
simulation experiments in well-defined model systems have also
played an essential role. This includes Brownian dynamics (BD)
\cite{heyeslodge} and molecular dynamics (MD) \cite{testardjcp}
investigat\textcolor{black}{ions} of the density- and temperature-dependence of the
liquid-gas phase separation kinetics in suddenly-quenched
Lennard-Jones \textcolor{black}{and square-well \cite{foffi0}} liquids, the  BD simulation of gel formation in the
colloidal version of the primitive model of electrolytes \textcolor{black}{
(a mixture of equally-sized oppositely charged
colloids) \cite{sanz}}, and MD  simulations of bigel formation driven by the
demixing of the two colloidal species of a binary
mixture\cite{foffi}. From the theoretical side, on the other hand,
a fundamental  framework is missing that provides an
unified description of both, the irreversible evolution of the
structure and dynamics of a liquid suddenly quenched to an
unstable homogeneous state, and the possibility that along this
irreversible process the system encounters the conditions for its
structural and dynamical arrest.

\textcolor{black}{Any theoretical attempt to model this complex phenomenon, however,  must be guided by the well-established experimental facts and simulation results above, according to which, arrested spinodal decomposition is indeed a possible mechanism for gelation in attractive colloidal suspensions. Although there is still an ongoing discussion regarding specific details, such as the location of the arrest line \cite{gibaud,luetalnature}, one can expect that some of these details are actually system-dependent, and that they will eventually be fully understood. In the meanwhile, it is important to focus on the simplest experimental model systems, where this phenomenology has been characterized as carefully and as thoroughly as possible. The simplest such model system should involve, for example, a perfectly monodisperse suspension of spherical particles with attractive interactions that do not require a third species (as colloid-polymer mixtures do), so that their interactions are state-independent, without the need to translate the real temperature into any form of effective temperature. The solutions of the globular protein lysozyme studied by Schurtenberger and collaborators \cite{cardinaux,gibaud} are, to our knowledge, the closest experimental realization of such an ideal model system. Thus, the experimental scenario established by this group may provide an invaluable guide that might serve as a reference for the initial theoretical advances in this field. For this reason, we now quote the main features of such experimental scenario.}

\textcolor{black}{According to Refs.  \cite{cardinaux,gibaud}, the experimental equilibrium phase diagram of aqueous lysozyme solutions shows  at high temperatures a stable gas-crystal equilibrium coexistence, and at lower temperatures, a metastable gas-liquid phase separation. Such equilibrium scenario is typical of systems of particles interacting through hard-sphere--like repulsions plus very short-ranged attractions. The non-equilibrium experiments performed consist of the fast temperature quench at fixed overall protein concentration (or volume fraction $\phi$), starting with the system in equilibrium at an initial temperature $T_i$ above the gas-liquid coexistence curve $T_{gl}(\phi)$, cooling the system to a final temperature $T$ lower than $T_{gl}(\phi)$. The region below the coexistence curve ``can be separated into three areas differing in their kinetic behavior: a region of complete demixing (I), gel formation through an arrested spinodal decomposition (II), and an homogeneous attractive glass (III)''  \cite{cardinaux,gibaud}. More specifically, for a shallow quench to a final temperature below the spinodal curve, $T\lesssim T_s(\phi)$, the classical sequence of  spinodal decomposition leading to complete phase separation is observed (region I). However, for quenches below a threshold (or ``tie-line'') temperature $T_0(\phi)$ ($< T_s(\phi)$), the spinodal domain structure initially coarsens but then completely arrests, forming gels via an arrested spinodal decomposition (region II). Microscopically, these gels correspond to a coexistence of the dilute fluid with a dense percolated glass phase. At lower temperatures, region II is delimited by the boundary with the glass phase (region III), where homogeneous attractive glasses can be reached by quenches at sufficiently low $T$. Many other observations, which can be consulted in the original references  \cite{cardinaux,gibaud}, enrich the observed experimental scenario with more detail, but the existence of these three kinetically distinct regions constitute the most relevant feature calling for a fundamental explanation. }

Some elements needed to construct the desired unifying theoretical framework may already
be contained in the classical theory of spinodal decomposition
\cite{cahnhilliard,cook,furukawa,langer,dhont}. Unfortunately, it
is not clear how to incorporate in these approaches the main
non-equilibrium features of the process of dynamic arrest,
including its history dependence and aging behavior. Similarly, it
does not seem obvious how to incorporate the non-stationary
evolution of the structure and dynamics of a liquid during the
process of spinodal decomposition in existing theories of glassy
behavior \cite{berthierreview}. For example, although many of the
predictions of conventional mode coupling theory (MCT)
\cite{goetze1,goetze2,goetze3,goetze4} have found beautiful
experimental verification, this theory is unable to describe irreversible
non-stationary processes. The reason for this is that, in its
present form, MCT is in reality a theory of the dynamics of
liquids in their thermodynamic equilibrium states. To overcome
this fundamental limitation, in 2000 Latz \cite{latz} proposed a
formal extension of MCT to situations far away from equilibrium
which, however, has not yet found a specific quantitative
application. \textcolor{black}{In this context, let us also refer to the mean-field theory of the aging dynamics of glassy spin systems developed by Cugliandolo and Kurchan \cite{cugliandolo1}. This theory has also made relevant detailed predictions that have been verified in experiments and simulations. Unfortunately, the models involved lack a geometric structure and hence cannot describe the spatial evolution
of real colloidal glass formers.}

\textcolor{black}{MCT's} limitation above is shared by the self-consistent generalized
Langevin equation (SCGLE) theory of dynamic arrest
\cite{rmf,todos1,todos2}, a theory that predicts, in a manner
completely analogous to MCT, the existence of dynamically arrested
states. In contrast to MCT, however, the SCGLE theory has been
extended to describe the irreversible non-equilibrium evolution of
glass-forming liquids, thus resulting in the
\emph{``non-equilibrium self-consistent generalized Langevin
equation''} (NE-SCGLE) theory \cite{nescgle0,nescgle1, nescgle2}.
This extended theory provides a conceptually simple picture of the
crossover from ergodic equilibration to non-equilibrium aging
\cite{nescgle3,nescgle4}, offering a perspective of the glass
transition in which the ``waiting''-time $t$ becomes a relevant
active variable and in which indeed the measured properties of the
system may depend on the protocol of preparation and on the
duration of the actual measurements. Thus, this non-equilibrium
theory may provide the general framework in which to build the
missing unified description of the processes of \emph{arrested}
spinodal decomposition.

The main purpose of the present paper is
to take the first steps in  \textcolor{black}{exploring this possibility. The first of these steps is to discuss the
manner in which the NE-SCGLE theory may be adapted to address the
description of arrested spinodal decomposition. The second refers to obtaining and presenting the
concrete results of its application to a specific
model system that permits at least qualitative contact with the observed behavior of an experimental model system such as that described above. The physical interpretation of our theoretical results, however, is not as straightforward as, for example, calculating the equilibrium properties of a system using century-old and well-established statistical thermodynamic methods (such as calculating the partition function \cite{mcquarrie}). Instead, because of the non-equilibrium and non-linear nature of the phenomenon described, the interpretation of the specific predictions is much more subtle, involving features that might seem at first sight strongly counterintuitive. Thus, the third and most relevant step, is to properly interpret these results. }

\textcolor{black}{Regarding the first of these steps, let us mention that the}  most general version of the NE-SCGLE theory of irreversible
processes in colloidal liquids \cite{nescgle0,nescgle1, nescgle2}
consists of two fundamental time-evolution equations, one for the
mean value $\overline{n}(\textbf{r},t)$, and another for the
covariance $\sigma(\textbf{r},\textbf{r}';t)\equiv
\overline{\delta n (\textbf{r},t)\delta n (\textbf{r}',t)}$ of the
fluctuations $\delta n(\textbf{r},t) = n(\textbf{r},t)-
\overline{n}(\textbf{r},t)$ of the local concentration profile
$n(\textbf{r},t)$ of a colloidal liquid. Thus, in the following
section \textcolor{black}{and in the appendix} we review this general description, to discuss
the form adopted by these equations in the absence of external fields,
and the possibility that they allow the description of a
fingerprint of spinodal decomposition, namely, the early stage of
the amplification of spatial heterogeneities. The resulting
equations are thus restricted to allow the discussion of
the simplest explicit protocol of thermal and mechanical
preparation, namely, an instantaneous isochoric quench, to
describe the spontaneous relaxation of the system toward its
equilibrium state or toward predictable arrested states. \textcolor{black}{To simplify the reading, Section \ref{sectionII} is essentially a summary of the specific NE-SCGLE equations that we shall actually solve, while the appendix contains a description of the main derivations and approximations leading to these specific equations. }

In Sect. \ref{sectionIII} we apply the resulting approximate
theory to the discussion of the possibility that the process of
spinodal decomposition may be interrupted by the emergence of
dynamic arrest conditions. For the sake of concreteness, there we
shall have in mind a very specific model system, namely, the
``hard-sphere plus attractive Yukawa'' (HSAY) potential, subjected
to an instantaneous quench at $t=0$ from an initial temperature
$T_i$ to a final temperature $T_f=T\ (< T_i)$, with the volume
fraction $\phi$ held fixed. The main focus of this application is
the determination of the dynamic arrest diagram in the region of
unstable homogeneous states. According to the predicted scenario,
the spinodal curve turns out to be the borderline between the
regions of ergodic and non-ergodic homogeneous states, and the
high-density liquid-glass transition line, whose high-temperature
limit corresponds to the well-known hard-sphere glass transition,
actually intersects the spinodal curve at lower temperatures and
densities, and continues inside the spinodal region as a
glass-glass transition line.

\textcolor{black}{Some elements of this predicted scenario, however, may seem strongly counterintuitive, and not consistent with experimental observations. For example, this scenario predicts that any quench inside the spinodal region will meet conditions for dynamic arrest, while we know that at least for shallow quenches the system will definitely phase-separate. A satisfactory understanding of these features requires a detailed discussion of a number of issues whose  subtlety derives from the lack of familiarity with the solutions of the full NE-SCGLE non-linear equations. Thus, in Sect. \ref{sectionIV} we discuss the physical interpretation of this theoretical scenario by analyzing in more detail the predicted long-time non-equilibrium arrested structure factor $S_a(k;\phi,T)$, particularly the length scale $\xi_a(\phi,T)=2\pi/k_{max}$ associated with the position $k_{max}$ of its small-$k$ peak. This length represents the size of the spatial heterogeneities typical of the early stage of spinodal decomposition described by the Cahn-Hilliard-Cook (CHC) classical linear theory of spinodal decomposition \cite{cahnhilliard,cook}, shown to be a particular limit of the NE-SCGLE theory. This analysis results in the recognition that indeed, shallow quenches lead to full phase separation whereas deeper quenches lead to the formation of gels. Thus, in Sect. \ref{sectionIII.6} we finally establish a more detailed connection between our theoretical scenario and the experimental observations in the lysozyme real model system, and  in Sect. \ref{sectionIV.3} we summarize the main results of this work.}

\section{Review of the NE-SCGLE theory.}\label{sectionII}

Let us start by considering a Brownian liquid  formed by $N$ particles that execute Brownian motion (characterized by a short-time self-diffusion coefficient $D_0$) in a volume $V$ while interacting between them through a generic pair potential $u(r)=u_{HS}(r)+u_{A}(r)$ that  is the sum of a hard-sphere (HS) term $u_{HS}(r)$ with HS diameter $\sigma$, plus an attractive tail $u_{A}(r)$.  Let us imagine that we subject this system to a prescribed protocol
of thermal and/or mechanical treatment, described by the chosen
temporal variation of the reservoir temperature $T(t)$, of the total volume $V(t)$, and of the applied external fields $\psi({\bf r},t)$. In principle, the ultimate goal of the NE-SCGLE theory is to predict the response of our system to each possible
protocol of thermal and mechanical manipulation. For simplicity let us assume that the thermal diffusivity of the system is sufficiently large that any
temperature gradient is dissipated almost instantly (compared with
the relaxation of chemical potential gradients), so that the
temperature field inside the system can be considered uniform, and instantly
equal to the temperature of the reservoir, $T({\bf r},t)=T(t)$.

Let us now imagine that we use these time-dependent fields and
constraints to drive the system to a prescribed (but arbitrary) initial state characterized by a given mean concentration profile $\overline{n}^{0}(\textbf{r})$ and
covariance $\sigma^{0}(k;\textbf{r})$, and we then fix the field, the
temperature, and the volume afterward,  $\psi({\bf r},t)=
\psi({\bf r})$,  $T(t)=T$, and $V(t)=V$ for $t>0$. The NE-SCGLE theory will then allow
us to predict the most fundamental and primary information that
characterizes the intrinsic properties of the system, namely, its
spontaneous behavior during the irreversible relaxation towards
its new expected equilibrium state. This is described in terms of the time-evolution of the non-equilibrium mean concentration profile $\overline{n}(\textbf{r},t)$ and
covariance $\sigma(k;\textbf{r},t)$ for  $t>0$ (see Eqs. (\ref{difeqdlpap}) and (\ref{relsigmadif2pap}) of the appendix).

In Ref. \cite{nescgle3} we started the systematic discussion of this topic in the context of a model system involving only purely repulsive interactions. For this, we restricted ourselves further to the simplest explicit protocol of thermal and
mechanical preparation of our system, namely, its
\emph{instantaneous} cooling or heating from an initial
temperature $T_i$ to a final temperature $T_f$, so that
$T(t)=T_i$ for $t\le 0$ and $T(t)=T_f$ for $t> 0$, under
\emph{isochoric} conditions, $V(t)=V$ (so that
$\overline{n}(t)=\overline{n}$) at all times, and in the absence of applied
external fields, $\psi({\bf r},t)= 0$. As explained in Ref. \cite{nescgle3}, the NE-SCGLE theory will then describe the spontaneous response of the system in terms only of the non-stationary but uniform covariance  $\sigma(k;t)$, now written as the non-equilibrium static structure factor $S(k;t)\equiv  \sigma(k;t)/\overline{n}$, whose time-evolution equation  for $t>0$ reads
\begin{equation}
\frac{\partial S(k;t)}{\partial t} = -2k^2 D^0
b(t)\overline{n}\mathcal{E}_f(k) \left[S(k;t)
-1/\overline{n}\mathcal{E}_f(k)\right]. \label{relsigmadif2pp}
\end{equation}
In this equation $\mathcal{E}_f(k)\equiv \mathcal{E}_h(k;\overline{n},T_f)$ is the Fourier transform of the  thermodynamic functional derivative
$\mathcal{E}[{\bf r},{\bf r}';n,T]\equiv \left[ {\delta \beta\mu
[{\bf r};n,T]}/{\delta n({\bf r}')}\right]$, evaluated at the
uniform density and temperature profiles $n(\textbf{r}) = n$ and
$T(\textbf{r}) = T$, in which case it can be written as
$\mathcal{E}[\textbf{r},\textbf{r}';n,T]=
\mathcal{E}_h(|\textbf{r}-\textbf{r}'|;n,T)=\delta
(\textbf{r}-\textbf{r}')/n-c(|\textbf{r}-\textbf{r}'|;n,T)$ or, in
Fourier space, as $\mathcal{E}_h(k;n, T)=1/n-c(k;n, T)$, where
$c(r;n,T)$ is the so-called direct correlation function.

The non-stationary, non-equilibrium mobility function $b(t)$ entering in Eq. (\ref{relsigmadif2pp}) above, is the most important kinetic
property. Within the NE-SCGLE theory this property is determined by (see Refs. \cite{nescgle1,nescgle2})
\begin{equation}
b(t)= [1+\int_0^{\infty} d\tau\Delta{\zeta}^*(\tau; t)]^{-1},
\label{bdt}
\end{equation}
with the $t$-evolving, $\tau$-dependent friction coefficient
$\Delta{\zeta}^*(\tau; t)$ given approximately by
\begin{equation}
\begin{split}
  \Delta \zeta^* (\tau; t)= \frac{D_0}{24 \pi
^{3}\overline{n}}
 \int d {\bf k}\ k^2 \left[\frac{ S(k;
t)-1}{S(k; t)}\right]^2  \\ \times F(k,\tau; t)F_S(k,\tau; t)
\end{split}
\label{dzdtquench}
\end{equation}
in terms of $S(k;t)$ itself and of the collective and
self non-equilibrium intermediate scattering functions $F(k,\tau;
t)$ and $F_S(k,\tau; t)$, whose memory-function equations are
written, in Laplace space, as
\begin{gather}\label{fluctquench}
 F(k,z; t) = \frac{S(k; t)}{z+\frac{k^2D^0 S^{-1}(k;
t)}{1+\lambda (k)\ \Delta \zeta^*(z; t)}},
\end{gather}
and
\begin{gather}\label{fluctsquench}
 F_S(k,z; t) = \frac{1}{z+\frac{k^2D^0 }{1+\lambda (k)\ \Delta
\zeta^*(z; t)}},
\end{gather}
with  $F(k,z; t)$, $F_S(k,z; t)$, and  $\Delta \zeta^*(z; t)$, being the corresponding  Laplace transforms (LT) and with $\lambda (k)$ being a phenomenological ``interpolating
function" \cite{todos2}, given by
\begin{equation}
\lambda (k)=1/[1+( k/k_{c}) ^{2}], \label{lambdadk}
\end{equation}
where the  phenomenological cut-off wave-vector $k_c$ depends on the system considered. \textcolor{black}{Since in this paper we are only interested in semi-quantitative trends, here we adopt the value $k_c=1.305 (2\pi/\sigma)=8.2/\sigma $, determined in a calibration procedure involving simulation data of the hard-spheres system  \cite{gabriel}}.

\section{Dynamic arrest diagram of a simple model system.}\label{sectionIII}

In this section we apply the previous general theory  to the
discussion of the possibility that the process of spinodal
decomposition may be interrupted by the emergence of dynamic
arrest conditions. For this, let us consider a generic liquid
formed by $N$ particles in a volume $V$ interacting through a
simple pair potential $u(r)=u_{HS}(r)+u_{A}(r)$, which is the sum
of a hard-sphere term $u_{HS}(r)$ plus an attractive tail
$u_{A}(r)$. \textcolor{black}{Although all the results that we shall discuss here are independent of the specific functional form of $u_{A}(r)$, f}or the sake of concreteness, in what follows we shall
have in mind a very specific model system, namely, the
``hard-sphere plus attractive Yukawa'' (HSAY) potential, defined
as
\begin{equation}
u(r)=
\begin{cases}
\infty, & r < \sigma; \\
-\epsilon \frac{\exp[-z(r/\sigma -1)]}{(r/\sigma )}, & r>\sigma,
\end{cases}
\label{yukawa}
\end{equation}
where $\sigma$ is the hard sphere diameter, $\epsilon$ the depth
of the attractive Yukawa well at contact,  \textcolor{black}{and $z^{-1}$ its
decay length (in units of $\sigma$)}. For given $\sigma$,
$\epsilon$, and $z$, the state space of this system is spanned by
two macroscopic variables, namely, the number density $n=N/V$ and
the temperature $T$, which we express in dimensionless form, as
$n^*\equiv n\sigma^3$ and $T^*\equiv k_BT/\epsilon$. In practice,
however, from now on we shall use $\sigma$ as the unit of length,
and $\epsilon/k_B$ as the unit of temperature, so that $n^*= n$
and $T^*=T$; most frequently, however, we shall refer to the
hard-sphere volume fraction $\phi\equiv \pi n/6$.

\subsection{Equilibrium thermodynamic properties.}\label{subsectionIII.1}

The most fundamental thermodynamic property of this system, that
we need to specify in order to apply the NE-SCGLE theory described
above, is the thermodynamic functional derivative
$\mathcal{E}[{\bf r},{\bf r}';n,T]\equiv \left[ {\delta \beta\mu
[{\bf r};n,T]}/{\delta n({\bf r}')}\right]$. Evaluated at the
uniform density and temperature profiles $n(\textbf{r}) = n$ and
$T(\textbf{r}) = T$, we have that
$\mathcal{E}[\textbf{r},\textbf{r}';n,T]=
\mathcal{E}_h(|\textbf{r}-\textbf{r}'|;n,T)=\delta
(\textbf{r}-\textbf{r}')/n-c(|\textbf{r}-\textbf{r}'|;n,T)$ or, in
Fourier space, as $\mathcal{E}_h(k;n, T)=1/n-c(k;n, T)$, with
$c(r;n,T)$ being  the direct correlation function. In the present
paper we shall rely on the approximate prescription proposed by
Sharma and Sharma \cite{sharmasharma} to determine this
thermodynamic property, which for the HSAY potential above reads
\begin{equation}
c(k;\phi, T) = c_{HS}(k;\phi) -\beta u_A(k), \label{sharma}
\end{equation}
in which $c_{HS}(k;\phi)$ is the FT of the direct correlation
function of the hard-sphere liquid, approximated by the
Percus-Yevick approximation with Verlet-Weis correction (PYVW)
\cite{percusyevick,verletweiss} and $\beta u_A(k)$ is the FT of
the attractive potential (in units of the thermal energy
$k_BT=1/\beta$), i.e., $\beta u_A(k)\equiv 4\pi \int_\sigma^\infty
dr r^2 [\sin (kr)/(kr)]\beta u_A(r)$. Thus, for the HSAY
potential, $\beta u_\sigma(k)\equiv -(4\pi/T^*) \int_1^\infty dr r^2
[\sin (kr)/(kr)]\exp[-z(r-1)]/r =  -(4\pi/T^*) \left[ (k \cos{k}+z \sin{k})/k(k^2+z^2)\right]$.

\begin{figure}
\begin{center}
\includegraphics[scale=.28]{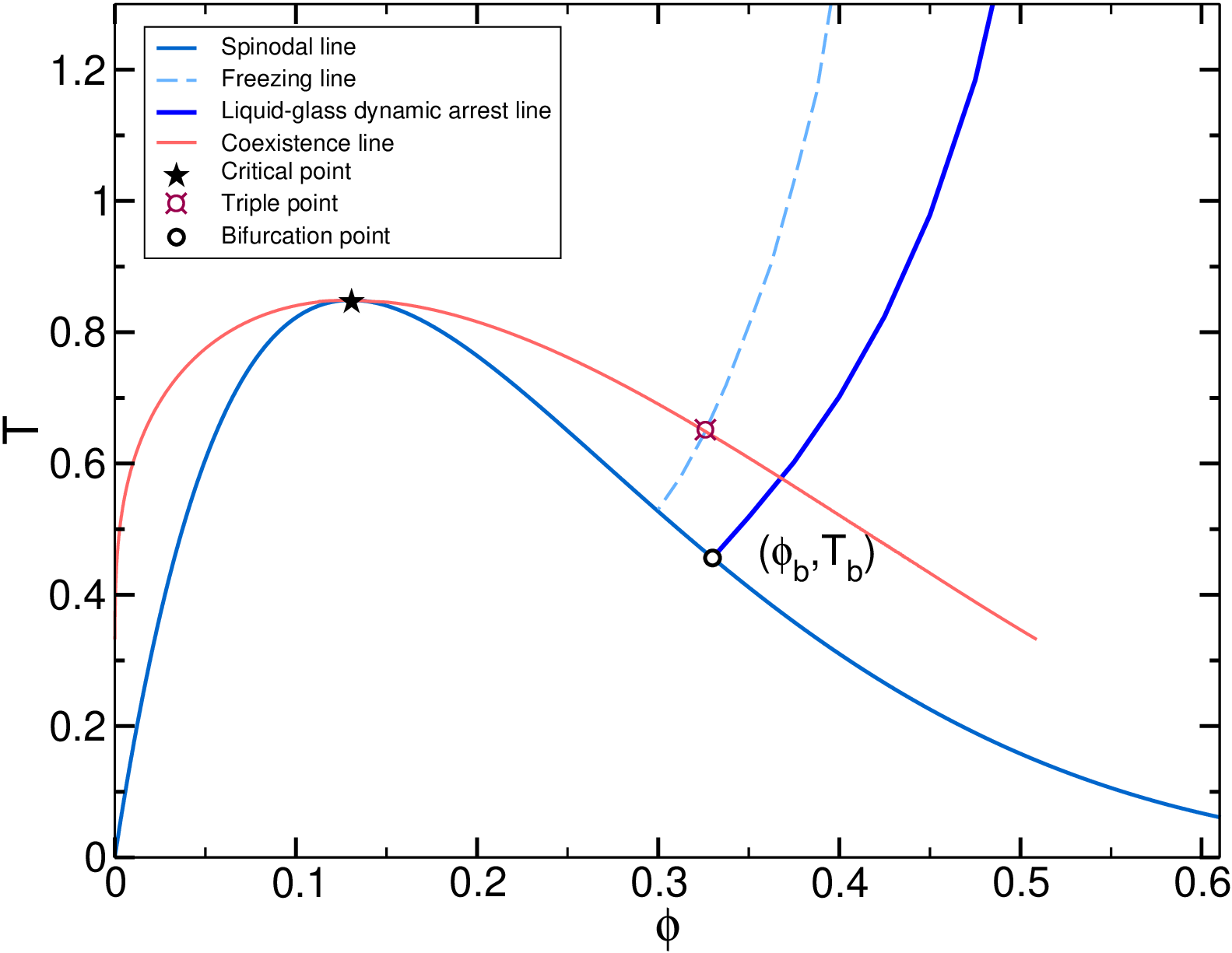}
\caption{\textcolor{black}{Binodal and spinodal lines of the gas-liquid coexistence of the HSAY model fluid in Eq. (\ref{yukawa})  with $z=2$, calculated with the Sharma-Sharma approximation for $\mathcal{E}_h(k;n, T)$. The corresponding critical point (black star) is located at $(\phi,T)=(0.13,0.85)$. Also shown is the freezing line, calculated from the Hansen-Verlet condition that the maximum of $S^{(eq)}(k;\phi, T)= 1/\overline{n}\mathcal{E}_h(k;\phi, T)$ reaches 2.85. The corresponding triple point (soft red circle) is located at $(\phi,T)=(0.33,0.65)$. Finally, the dark blue solid line is the liquid-glass dynamic arrest line, calculated from the solution $\gamma^{eq}(T, \phi)$ of the
equilibrium ``bifurcation equation'', Eq. (\ref{nep5ppequil}). The dark circle denotes the intersection $(\phi_b, T_b) =(0.33,0.45)$ of this arrest line with the spinodal curve. }}\label{fig1}
\end{center}
\end{figure}

For fixed $z$, the equilibrium phase diagram of the HSAY model
system in the state space ($\phi, T$) contains the gas, liquid and \textcolor{black}{(crystalline) solid phases. The previous approximation  for $\mathcal{E}_h(k;n, T)$ allows us to sketch the most prominent features of the fluid phases. For example, the gas-liquid transition involves a coexistence region, with its associated binodal and spinodal lines. The former can be determined by integration of $\beta [\partial p/ \partial n]_T=\mathcal{E}_h(k=0;\phi, T)$ to get the equation of state, along with Maxwell's construction. The latter can be determined from the condition for  thermodynamic instability of uniform fluid states, namely, $\mathcal{E}_h(k=0;n, T)\le 0$. The resulting  binodal and spinodal lines are illustrated in Fig. \ref{fig1} for the HSAY system with $z=2$. On the other hand, the \emph{equilibrium} static structure factor of the fluid state is determined by the equilibrium (Ornstein-Zernike) condition $S^{(eq)}(k;\phi, T)=
1/\overline{n}\mathcal{E}_h(k;\phi, T)$. From $S^{(eq)}(k;\phi,
T)$   the freezing line, which is another boundary of stability of the uniform liquid state, can be readily sketched
using the phenomenological Hansen-Verlet condition \cite{hansenverlet} that the height  $S^{(eq)}_{max}(\phi, T)$ of the main peak of the equilibrium static structure factor $S^{(eq)}(k;\phi, T)$ reaches  the value  $S^{(eq)}_{max}(\phi, T)=2.85$. Besides the  spinodal, binodal and freezing lines, Fig. \ref{fig1} also exhibits the corresponding gas-liquid critical point, located at $(\phi,T)=(0.13,0.85)$, and the triple point at $(\phi,T)=(0.33,0.65)$. These equilibrium lines serve as a reference to the introduction, in the following subsection,  of the liquid-glass dynamic arrest line, also obtained from $S^{(eq)}(k;\phi, T)= 1/\overline{n}\mathcal{E}_h(k;\phi, T)$.}

\subsection{Dynamic arrest diagram from the equilibrium SCGLE theory.}\label{subsectionIII.2}

The first relevant task in the description of dynamic arrest
phenomena in a given system, such as that in our example, is the
determination of the dynamic arrest diagram, i.e., the region
containing the state points $(\phi, T)$ where the system will be
able to reach thermodynamic equilibrium (ergodic region) and the
region where the system, prevented from crystallizing,  will be trapped in a dynamically arrested
state (non-ergodic region). Under normal circumstances this can be
approached solving the so-called bifurcation equations of mode
coupling theory \cite{goetze1}, or the corresponding equations of
the \emph{equilibrium} version of the SCGLE theory \cite{todos2}.
The latter consists of the following equation for the equilibrium
localization length squared $\gamma^{(eq)}(\phi, T)$,
\begin{equation} \frac{1}{\gamma^{(eq)}} =
\frac{1}{6\pi^{2}\overline{n}}\int_{0}^{\infty }
dkk^4\frac{\left[S^{(eq)}(k;\phi, T)-1\right] ^{2}\lambda^2
(k)}{\left[\lambda (k)S^{(eq)}(k;\phi, T) +
k^2\gamma^{(eq)}\right]\left[\lambda (k) +
k^2\gamma^{(eq)}\right]}. \label{nep5ppequil}
\end{equation}
The only input of this equation is the \emph{equilibrium} static
structure factor $S^{(eq)}(k;\phi, T)=
1/\overline{n}\mathcal{E}_h(k;\phi, T)$ at a given state point
$(\phi, T)$. If the solution is $\gamma^{(eq)}(\phi, T)=\infty$,
we conclude that the state point $(\phi, T)$ lies in the ergodic
region, whereas if $\gamma^{(eq)}(\phi, T)$ is finite, the point
$(\phi, T)$ lies in the non-ergodic region of state space. In this
manner we can determine the dynamic arrest transition line.

This procedure has been performed in the past for several model
systems, for which the equilibrium static structure factor
$S^{(eq)}(k;\phi, T)$, given in terms of the equilibrium condition
$S^{(eq)}(k;\phi, T)= 1/\overline{n}\mathcal{E}_h(k;\phi, T)$, is
well-defined in the entire state space $(\phi, T)$. In the present
case we have carried out this exercise for our HSAY model, and the
result is the liquid-glass dynamic arrest transition line
$(\phi_c, T_c)$ represented in Fig. \ref{fig1} by the dark (blue)
solid line, with the region to the left and above this curve
corresponding to ergodic states. This dynamic arrest line extends
from its high temperature limit, corresponding to the hard-sphere
glass transition at $(\phi, T)=(0.582,\infty)$, down to its
intersection $(\phi_b, T_b) =(0.33,0.45)$  with the spinodal
curve, indicated by the empty circle in the figure. Unfortunately,
for systems with thermodynamically unstable (spinodal) regions,
like our HSAY model, we can only apply this method outside such
unstable region, since for the state points $(\phi, T)$
\emph{inside} the spinodal no equilibrium static structure factor
$S^{(eq)}(k;\phi, T)$ exists that corresponds to spatially uniform
states. It is at this point that the power of the
\emph{non-equilibrium} version of the SCGLE theory is manifested,
since this general theory does provide a proper manner to overcome
this limitation of the equilibrium theory thus offering an
intriguing and unexpected qualitative scenario of the interplay
between dynamic arrest and gas-liquid phase separation. As we
shall see below, the alternative method is NOT based on the use of
the equilibrium bifurcation equation, Eq. (\ref{nep5ppequil}), but
on its non-equilibrium generalization, Eq. (\ref{nep5ppdu}) below.

\subsection{Detecting dynamic arrest transitions using the
\emph{NE}-SCGLE theory.}\label{subsectionIII.3}

\textcolor{black}{Let us now use the NE-SCGLE theory to detect dynamic arrest transitions.
We start by noticing that from the very structure of Eq.
(\ref{relsigmadif2pp}) one can infer the existence of two
fundamentally different kinds of stationary solutions,
representing two fundamentally different kinds of stationary
states of matter. The first corresponds to the condition
in which stationarity is attained because the factor $\left[S(k;t)
-1/\overline{n}\mathcal{E}_f(k)\right]$ on the right side of Eq.
(\ref{relsigmadif2pp}) vanishes, i.e., because $S(k;t)$ is able to
reach its thermodynamic equilibrium value
$S^{(eq)}(k;\overline{n},T_f)=1/\overline{n}\mathcal{E}_h(k;\overline{n},T_f)$,
while the mobility $b(t)$ attains a \emph{finite} positive
long-time limit $b_f^{eq}$. These stationary solutions correspond, of course, to
the ordinary thermodynamic equilibrium states, in which the difference
$[S(k;t)-1/\overline{n}\mathcal{E}_f(k)]$ decays (according
to Ec. (\ref{relsigmadif2pp}), and in a gross approximation) exponentially
fast, $[S(k;t)-1/\overline{n}\mathcal{E}_f(k)]\propto \exp[-
t/t^{eq}]$, with the \emph{equilibration} time $t^{eq}$ estimated
as $t^{eq}(\overline{n},T_f) \propto
\left[2k_{max}^2D^0\overline{n}\mathcal{E}_f(k_{max})b_f^{eq}
\right]^{-1}$.}

 \textcolor{black}{The second class of stationary solutions of Eq.
(\ref{relsigmadif2pp}) emerges from the possibility that the
long-time asymptotic limit of the kinetic factor $b(t)$ vanishes,
so that $dS(k;t)/dt$ vanishes at long times without requiring the
equilibrium condition $\left[S(k;t)
-1/\overline{n}\mathcal{E}_h(k;\overline{n},T_f)\right]=0$ to be
fulfilled. This mathematical possibility has profound and general
implications. For example, under these conditions  $S(k;t)$ will
now approach a distinct non-equilibrium stationary limit, denoted
by $S_a(k)$, which is definitely different from the expected
equilibrium value $S^{(eq)}(k;\overline{n},T_f)=
1/\overline{n}\mathcal{E}_h(k;\overline{n},T_f)$. Furthermore, the
difference $[S(k;t)-S_a(k)]$ is predicted to decay to zero not
exponentially, but in a much slower fashion, namely, as
$t^{-x}$, with an exponent $x$ near unity \cite{nescgle2, nescgle3}.
Contrary to ordinary equilibrium states, whose properties are
determined solely by the fundamental condition of maximum entropy,
the properties of these stationary but
intrinsically non-equilibrium states, such as $S_a(k)$, may depend on the
preparation protocol (in our example of the instantaneous isochoric quench, on $T_i$ and $T_f$).}

\textcolor{black}{For a given instantaneous isochoric quench from an initial
temperature $T_i$  to a lower final temperature $T_f$, one can also predict
if the system will equilibrate or will be trapped in a non-equilibrium arrested
state by analyzing the stationary solutions of  Eq. (\ref{relsigmadif2pp}). This analysis
simplifies greatly with the change of
variables from the actual evolution time $t$ to the so-called
``material'' time $u(t)$ defined by \cite{nescgle3}
\begin{equation}
u(t) \equiv \int_0^t b(t')dt'. \label{udt}
\end{equation}
This allows us to write, for example, the actual solution $S(k;t)$
of Eq. (\ref{relsigmadif2pp}) as $S(k;t)= S^*(k;u(t))$, with the
function $S^*(k;u)$ being the solution of
\begin{equation} \frac{\partial
S^*(k;u)}{\partial u} = -2k^2 D^0 \overline{n}\mathcal{E}_f(k)
\left[S^*(k;u) -1/\overline{n}\mathcal{E}_f(k)\right]
\label{relsigmadif2ppu}
\end{equation}
i.e.,
\begin{equation}
S(k;t)= S^*(k;u(t)) \equiv
[\overline{n}\mathcal{E}_f(k)]^{-1}+\left \{ S_i(k)-
[\overline{n}\mathcal{E}_f(k)]^{-1}\right \}e^{-2k^2D_0
\overline{n}\mathcal{E}_f(k)u(t)}, \label{solsigmadkt}
\end{equation}
where $S_i(k)\equiv S(k;t=0)=S^*(k;u=0)$ represents the
(arbitrary) initial condition.}

\textcolor{black}{This expression  interpolates $S^*(k;u)$ between its initial value
$S_i(k)$ and the value of the thermodynamic property
$[\overline{n}\mathcal{E}_f(k)]^{-1}$. Thus, if nothing impedes
the system from reaching equilibrium at the state point
$(\overline{n},T_f)$, the non-stationary static structure factor
$S(k;t)$ will eventually attain its equilibrium value
$S^{(eq)}(k;\overline{n},T_f)= [\overline{n}\mathcal{E}_f(k)]
^{-1}$. This, however, is only one of the two possibilities
described above. In order to determine which of them will govern
the course of the spontaneous response of our system, we actually
input $S^*(k;u)$ in the ``bifurcation'' equation for the square
localization length $\gamma(t)$ at evolution time $t$ (i.e., the
long-$\tau$ asymptotic value of the mean squared displacement).
Denoting $\gamma(t)=\gamma^*(u(t))$ simply as $\gamma^*(u)$, such
an equation reads
\begin{equation}
\frac{1}{\gamma^*(u)} =
\frac{1}{6\pi^{2}\overline{n}}\int_{0}^{\infty }
dkk^4\frac{\left[S^*(k;u)-1\right] ^{2}\lambda^2
(k;u)}{\left[\lambda (k;u)S^*(k;u) +
k^2\gamma^*(u)\right]\left[\lambda (k;u) + k^2\gamma^*(u)\right]}.
\label{nep5ppdu}
\end{equation}
As explained in Ref. \cite{nescgle3}, if we find that
$\gamma^*(u)=\infty$ for $0\le u \le \infty$, we conclude that the
system will be able to equilibrate after this quench, and hence,
that the point $(\overline{n},T_f)$ lies in the ergodic region.}
If, instead, a finite value $u_a$ of the parameter $u$ exists,
such that $\gamma^*(u)$ remains infinite \emph{only} within a
finite interval $0\le u < u_a$, the system will no longer
equilibrate, but will become kinetically arrested,  with the non-equilibrium asymptotic
structure factor $S_a(k)$ given by
\begin{equation}
S_a(k)\equiv S^*(k;u_a) =
[\overline{n}\mathcal{E}_f(k)]^{-1}+\left \{ S_i(k)-
[\overline{n}\mathcal{E}_f(k)]^{-1}\right \}e^{-2k^2D_0
\overline{n}\mathcal{E}_f(k)u_a}.  \label{solsigmadkta}
\end{equation}
\textcolor{black}{The parameter $u_a$ and the corresponding long-time
asymptotic square localization length  $\gamma^*_a \equiv \gamma^*(u_a)$
are dynamic order parameters in the sense that if $u_a=\gamma_a^*=\infty$,
the system will reach equilibrium, but if $u_a$ and $\gamma_a^*$ are finite,
the system will be dynamically arrested. Both of them, as well as the non-equilibrium
static structure factor $S_a(k)$, will depend on the protocol of the quench (in our
case, on $\phi$, $T_i$, and the final temperature $T_f$, which from now on will be
denoted simply as $T$). Thus, if one determines the functions $u_a=u_a(\phi, T_i,T)$,
$\gamma_a^*=\gamma_a^*(\phi, T_i,T)$, and $S_a(k)=S_a(k;\phi, T_i,T)$, one
can in principle draw the dynamic arrest diagram.}

\begin{figure}
\begin{center}
\includegraphics[scale=.28]{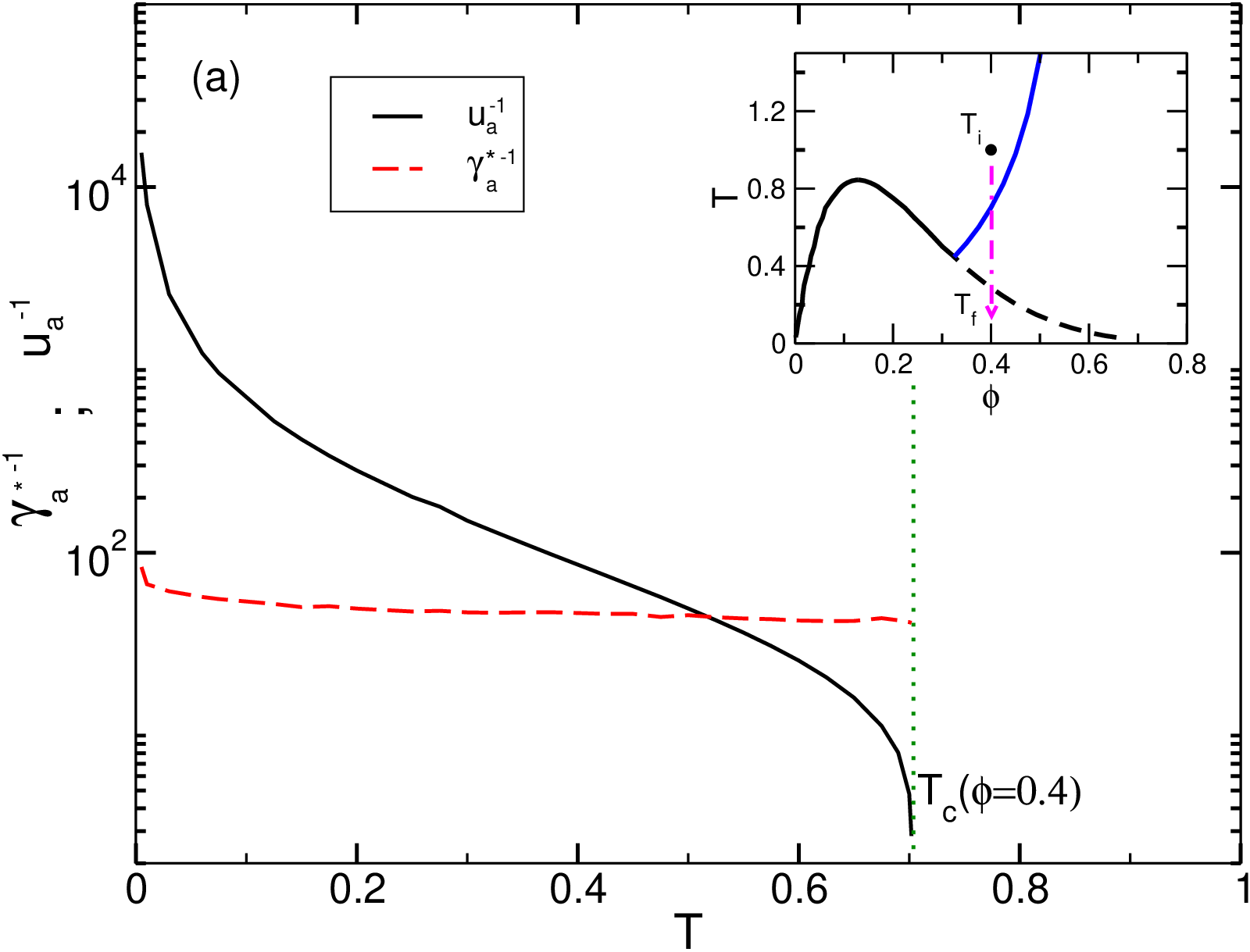}\hskip0.5cm
\includegraphics[scale=.28]{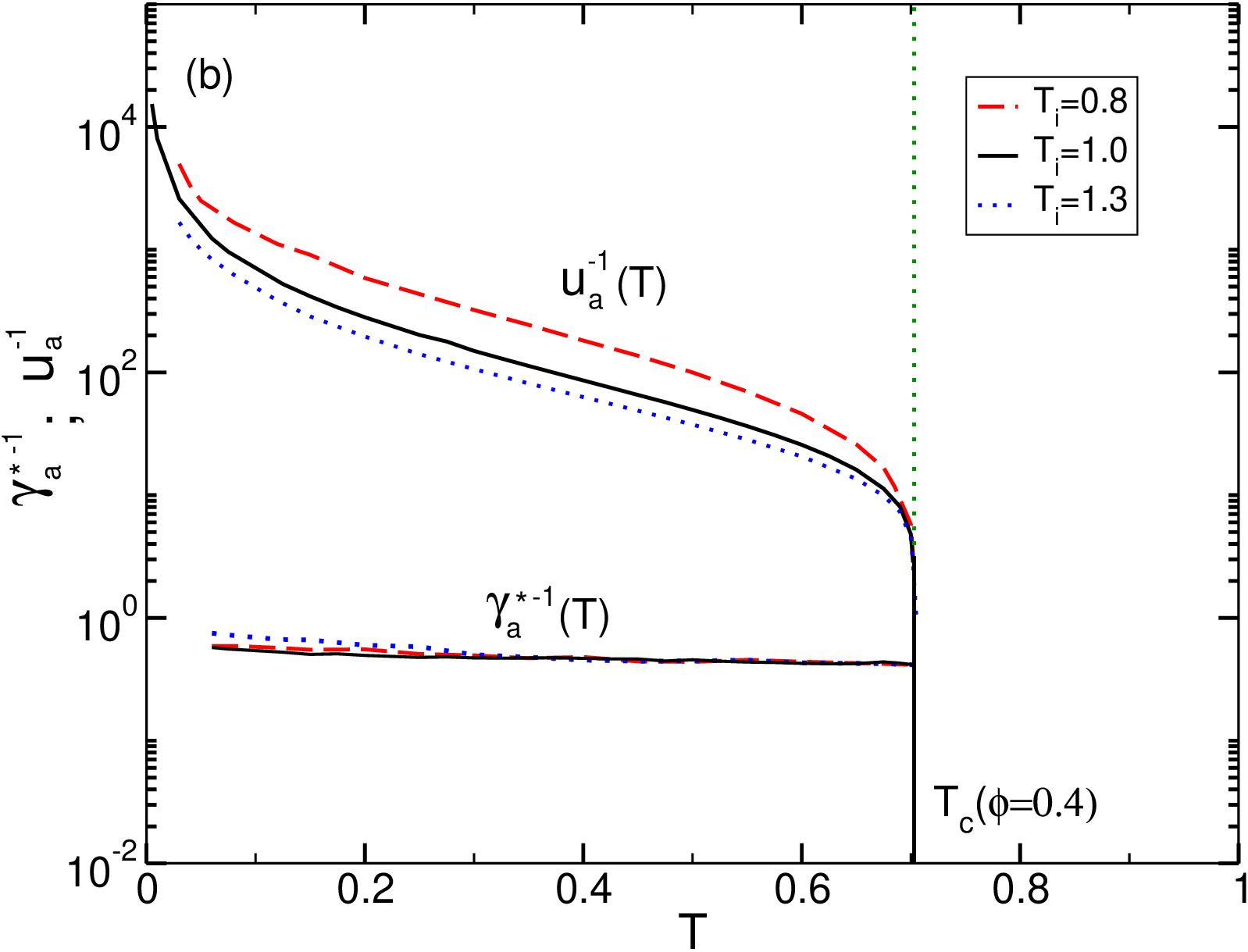}
\caption{(a) Dependence of $u_a(T)$ (solid line) and of
$\gamma^*_a(T)$ (dashed line) on the final temperature $T$ of an
instantaneous quench with initial temperature $T_i=1.0$ and fixed
volume fraction $\phi=0.4$. The inset reproduces essentially Fig.
\ref{fig1}, with the arrow indicating this illustrative isochoric
quench. (b) $u_a(T)$ for the previous quench and for other two
quenches that start at different initial temperatures, namely,
$T_i=1.3$ and with $T_i=0.8$. Notice that the three cases yield
the same result for the transition temperature, $T_c(\phi=0.4) =
0.7$.} \label{fig2}
\end{center}
\end{figure}

We have implemented this protocol and here we present the results
of its application to the concrete illustrative case involving the
HSAY model. Let us first consider isochores with $\phi> \phi_b$,
which lie to the right of the intersection indicated by the dark circle
in Fig. \ref{fig1}. A representative isochoric quench of this
class is schematically indicated in the inset of Fig.
\ref{fig2}(a) by the downward arrow along the isochore $\phi=0.4$.
In the main panel of this figure we plot the resulting $u^{-1}_a(T)$ as
a function of $T$ (solid line), together with the corresponding
value of $\gamma^{*-1}_a(T)$ (dashed line).

The first feature to notice is that indeed, for quenches to final
temperature $T$ above the critical temperature $T_c=0.702$,
$u_a(T)$ is always infinite (so that $u_a^{-1}(T)=0$, and hence,
do not appear in the logarithmic  window of the figure), whereas for
$T\le T_c$ we detect a finite value of $u_a(T)$. This value,
however, diverges as $T$ approaches $T_c$ from below, and
decreases monotonically as the final temperature $T$ decreases.
According to the results in Fig. \ref{fig2}(a), for $T\le T_c$
also $\gamma^{*}_a(T)$ attains finite values, thus reaffirming the
existence of an arrested phase in this temperature regime. In
fact, these results for $\gamma^*_a(T)$ reveal a second relevant
feature, namely, that this dynamic order parameter changes
discontinuously at $T=T_c$, from a value $\gamma^{*-1}_a(T_c^-)=41$ to
a value $\gamma^{*-1}_a(T_c^+)=0$. We notice that this jump in
the value of  $\gamma^*_a(T)$ corresponds to the jump in
the solution $\gamma^{eq}(T)$ of the equilibrium bifurcation
equation, Eq. (\ref{nep5ppequil}), and it occurs at the same
critical temperature $T_c=0.702$, thus confirming that both
methods refer to the same dynamic arrest transition, and that this
is a ``type B'' dynamic arrest transition in the language of
mode coupling theory. To complete this identification, we repeated
this procedure at other volume fractions to determine the
$\phi$-dependence of the dynamic arrest transition temperature
$T_c(\phi)$, thus finding that indeed, outside the spinodal curve
(i.e., for volume fractions $\phi$ larger than the volume fraction
of the intersection point $(\phi_b, T_b) =(0.33,0.45)$) this
method leads to exactly the same liquid-glass dynamic arrest line
previously obtained using Eq. (\ref{nep5ppequil}).

Let us notice, however, that the curves describing  the dependence
of  $u_a(T)$ on the final temperature $T$ do depend on the initial
temperature $T_i$ of the quench. However, the resulting transition
temperature, $T_c(\phi=0.4) = 0.702$, is independent of $T_i$.
This is illustrated in Fig. \ref{fig2}(b), which compares the
function $u_a(T)$ obtained with $T_i=1.0$ with the results for
$u_a(T)$ obtained with $T_i=1.3$ and $T_i=0.8$. This independence
of $T_c(\phi)$ on the initial temperature $T_i$ is also a
necessary condition for the present method to be a reliable
approach to the determination of the liquid-glass dynamic arrest
transition. In  Fig. \ref{fig2}(b) we have also included the
results for $\gamma^{*-1}_a(T)$ corresponding to the quench processes
with different initial temperatures, to show that $\gamma^*_a(T)$
is much less sensitive to the initial temperature of the quench.
Let us also notice  that for $\phi > \phi_b$, the results for
$\gamma^*_a(T)$ and $u_a(T)$ as a function of the final
temperature $T$, do not exhibit any discontinuous behavior when
crossing the spinodal line, i.e., their dependence on $T$ does not
detect the spinodal.  Thus, for $\phi > \phi_b$ the spinodal
condition $\mathcal{E}_h(k=0;\phi,T)=0$ does not seem to have any
significant effect on the dynamics of the system. \textcolor{black}{As discussed below, this rather astonishing result is a consequence of the structure of the expression for $S_a(k)$ in Eq. (\ref{solsigmadkta}), and} is in dramatic contrast with what happens in the complementary
regime, i.e., for volume fractions $\phi \le \phi_b$, as we \textcolor{black}{now}
explain.

\subsection{Dynamic arrest inside the spinodal
region.}\label{subsectionIII.3}

\begin{figure}
\begin{center}
\includegraphics[scale=.27]{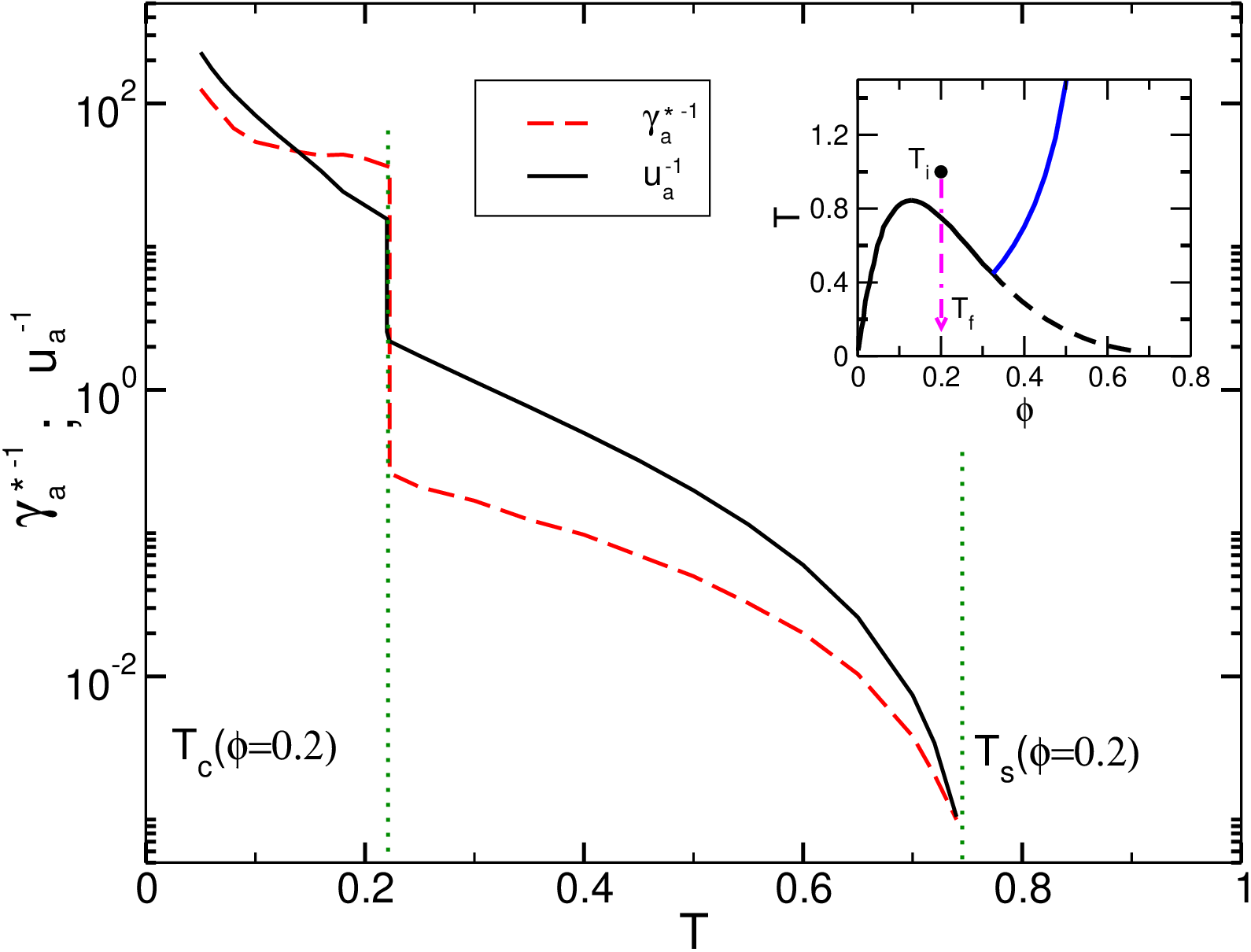} 
\caption{Dependence of $u_a(T)$ (solid line) and of
$\gamma^*_a(T)$ (dashed line) on the final temperature $T$ of an
instantaneous quench with initial temperature $T_i=1.0$ and fixed
volume fraction $\phi=0.2$. The inset reproduces essentially Fig.
\ref{fig1}, with the arrow indicating this illustrative isochoric
quench. \textcolor{black}{Notice that  $\gamma_a$ and $u_a$
are discontinuous at a temperature $T_{c}$, which implies the existence of a type-B glass-glass transition, whereas as T approaches to $T_s$  $\gamma_a$ and $u_a$ diverge, thus implying a type-A dynamic arrest transition at the spinodal.}} \label{fig3}
\end{center}
\end{figure}

The main advantage of the NE-SCGLE methodology just explained, for
determining the dynamic arrest transition line, is that its use
can be extended to the interior of the spinodal region. This
\textcolor{black}{sub}section continues the description of the use of this methodology,
but now to unveil the structure of the dynamic arrest diagram of
our model in the region inside the spinodal line. Thus, let us
repeat the calculations illustrated in Fig. \ref{fig2}, but now
for isochores corresponding to volume fractions  $\phi$ smaller
than the volume fraction $\phi_b=0.33$ of the intersection between
the liquid-glass dynamic arrest transition and the spinodal curve.
The corresponding illustrative example is contained in Fig.
\ref{fig3}, which plots the functions $u_a(T)$  and $\gamma^*_a
(T)$ as a function of $T$ for the HSAY model system subjected to
an instantaneous quench from an initial temperature $T_i=1.0$ to a
final temperature $T_f< 1.0$ at fixed $\phi=0.2$ (downward arrow in
the inset of the figure).

In contrast with the results for the isochore $\phi=0.4$, in Fig.
\ref{fig2}, we notice that in the present case  the functions
$u_a(T)$  and $\gamma^*_a (T)$ remain infinite for all
temperatures above a critical temperature $T_s=0.76$, and that
this singular temperature coincides precisely with the temperature
of the spinodal curve for that isochore. The results in Fig.
\ref{fig3} also indicate that both parameters, $u_a(T)$  and
$\gamma^*_a (T)$, are finite for $T < T_s$, and both diverge as
$T$ approaches $T_s$ from below. This reveals a rather dramatic
and unexpected conclusion, namely, that the spinodal curve,
besides being the threshold of thermodynamic instability, now
turns out to be also the threshold of non-ergodicity. Furthermore,
the fact that $\gamma^*_a (T)$ diverges as $T$ approaches $T_s$
from below determines that this transition from ergodic to
non-ergodic states occurs in a continuous fashion, i.e., that it
is a ``type A'' dynamic arrest transition in MCT language.

\begin{figure}
\begin{center}
\includegraphics[scale=.27]{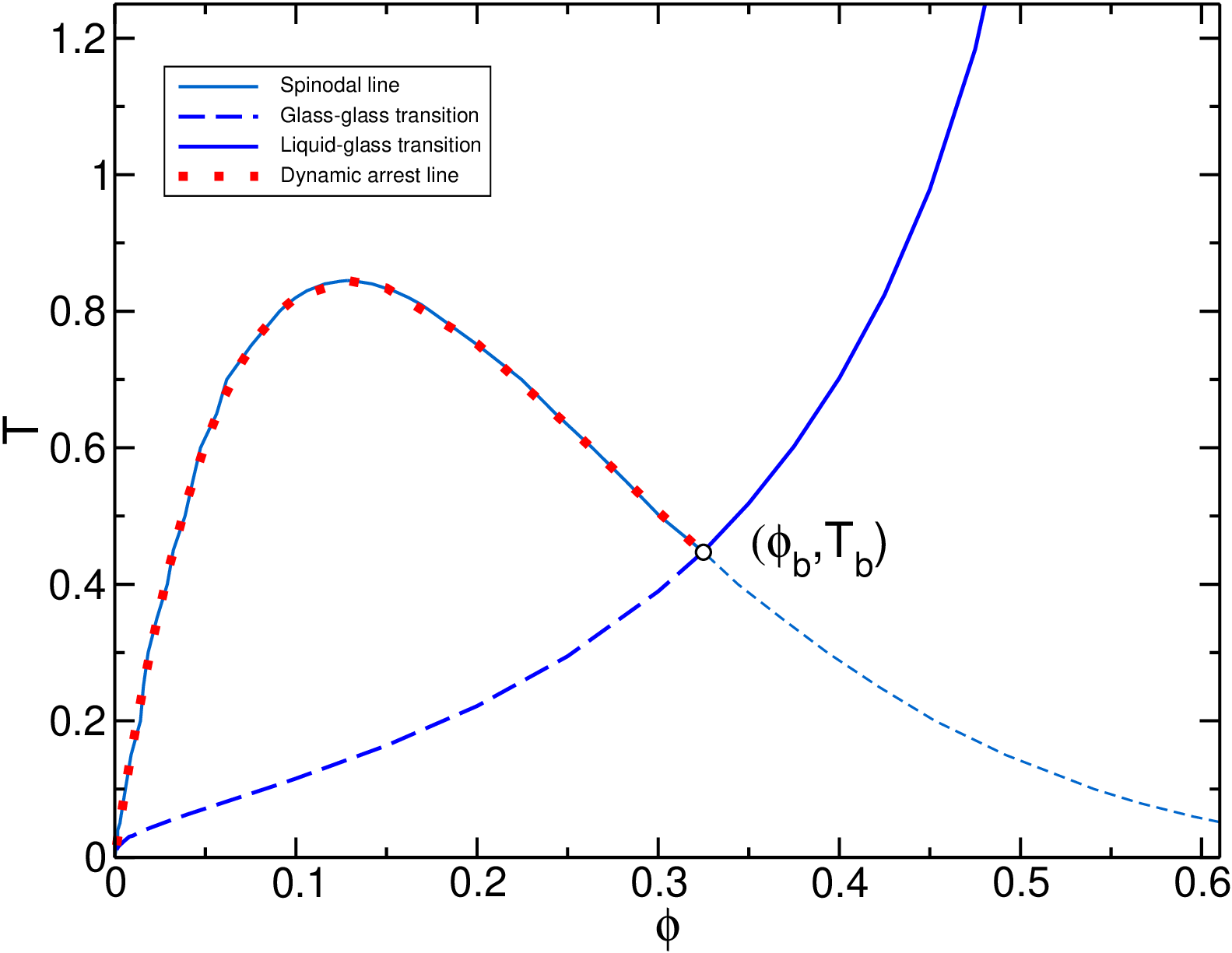}
\caption{ Dynamic arrest diagram exhibiting the low-$T$ (type B)
glass-glass transition (dark blue dashed curve), which continues
the liquid-glass transition (dark blue solid curve) to the
interior of the spinodal region. The (red) dotted line that
superimposes on the spinodal curve is a continuous (type A)
ergodic--nonergodic transition. These two transition\textcolor{black}{s} merge at the
bifurcation point $(\phi_b, T_b)$. \textcolor{black}{The soft blue dashed curve to the right of the bifurcation point  $(\phi_b, T_b)$ is the dynamically irrelevant part of the spinodal curve (see Subsection \ref{subsectionIII.5}).}} \label{fig3b}
\end{center}
\end{figure}

Examining again the same results in Fig.  \ref{fig3},   but now at
temperatures well below the spinodal curve, we see that the
parameters $u_a(T)$  and $\gamma^*_a (T)$ exhibit a discontinuity
at a lower temperature $T_c(\phi)$. This discontinuity reveals the
existence of still a second dynamic arrest transition, now
corresponding to a glass-glass ``type B'' transition, in which the
dynamic order parameter $\gamma^*_a (T)$  changes discontinuously
by about one order of magnitude, from a value
$\gamma^*_a(T_c^-)=36$ to another finite value
$\gamma^*_a(T_c^+)=4$. Notice that the value $\gamma^*_a (T^-)$ is
similar to the corresponding value at the isochore $\phi=0.4$.
Repeating these calculations at other isochores, we determine both
transition temperatures, $T_c(\phi)$ and $T_s(\phi)$, as a
function of volume fraction. The corresponding dynamic arrest
transition lines are presented in Fig. \ref{fig3b}. This figure
reveals a rather unexpected global scenario, in which (a) this
low-$T$ (type B) glass-glass transition (dark blue dashed curve)
turns out to be just the continuation to the interior of the
spinodal region, of the liquid-glass transition previously
determined outside the spinodal (dark blue solid curve), (b)  the
spinodal line itself is a continuous (type A) ergodic--nonergodic
transition, (c) these two transition\textcolor{black}{s} merge at the bifurcation
point $(\phi_b, T_b)$ to continue outside the spinodal as the
liquid-glass transition line that terminates at the hard-sphere
glass transition point $(\phi, T)=(0.582,\infty)$\textcolor{black}{, and (d) for $\phi > \phi_b$ the spinodal line does not have any apparent dynamic significance.}

\textcolor{black}{At first glance some of these predictions might appear counterintuitive. Let us remind, however, that they are based only on the analysis of the dependence of the dynamic order parameter $\gamma^*_a (\phi,T)$ on the final temperature $T$ of the instantaneous isochoric quenches analyzed. This order parameter is a functional of the long-time asymptotic value $S_a(k)$ of the non-stationary static structure factor $S(k;t)$. Thus, the analysis of the dependence of  $S_a(k;\phi,T)$ on $\phi$ and $T$ should provide additional pieces of structural information, while the waiting-time dependence of  $S(k;t)$ (and of all the other structural and dynamic properties) will provide valuable pieces of kinetic information. When these pieces are assembled together in a structural and kinetic jigsaw puzzle, one should expect that a sound and convincing scenario will emerge. For this, however, we must first identify and describe the most relevant of these pieces. The dynamic arrest diagram in Fig. \ref{fig3b} constitutes the initial and most basic piece for this discussion, but in the rest of the paper we shall identify others, equally fundamental.  }

\section{Physical interpretation of the dynamic arrest diagram.}\label{sectionIV}

The scenario described by the dynamic arrest diagram in Fig.
\ref{fig3b} is highly provoking, but also at first sight difficult
to reconcile with experience. First, we all know that simple
fluids, when quenched to the inside of the spinodal region, are expected to
fully phase-separate into the gas phase and its corresponding condensed
(liquid or crystal) phase. \textcolor{black}{ On the other hand, as explained in the introduction, there are numerous experimental and numerical evidences that} for
sufficiently deep quenches, the process of spinodal decomposition
may be interrupted by its interference with the dynamic arrest of
the condensed phase. Thus, the dynamic arrest diagram in Fig.
\ref{fig3b} may contain the fundamental explanation of this
experimental double scenario of full vs. arrested spinodal
decomposition.

To test this expectation, however, we must first
understand what exactly is the detailed physical meaning of these
NE-SCGLE predictions and what are their most relevant and
verifiable physical consequences. \textcolor{black}{In this section we discuss several pieces of information that derive from the solution of the NE-SCGLE equations, and which will dissipate some of the most subtle puzzles of the interpretation of this  dynamic arrest diagram. With these details taken into account, we shall see that the resulting theoretical scenario is actually quite consistent with the main features of the experimental scenario of arrested spinodal decomposition in lysozyme protein solutions, thus establishing a concrete link of our theory with physical reality.}

\subsection{\textcolor{black}{$S(k;t)$, $S^{(eq)}(k;\phi, T)$, and $S_a(k)$ below the spinodal line.}}\label{subsectionIV.1}

\textcolor{black}{In this subsection, for example, we discuss  the mechanism that allows the existence of a well-behaved non-equilibrium $S(k;t)$ for quenches to  state points $(\phi,T)$ inside the spinodal region, where the uniform fluid state has become thermodynamically unstable and the equilibrium structure factor $S^{(eq)}(k;\phi, T)$ does not exist. According to Eqs. (\ref{solsigmadkt}) and (\ref{solsigmadkta}), the
calculation of $S(k;t)$ and $S_a(k)$ requires the value of the thermodynamic property
$\mathcal{E}_h(k;\phi,T)$ for $k\ge 0$. As we know, the condition
$\mathcal{E}_h(k=0;\phi,T)> 0$ is a condition for the stability of
uniform thermodynamic states, a condition held for all
temperatures $T$ above the spinodal temperature $T_s(\phi)$. Thus,
the spinodal condition $\mathcal{E}_h(k=0;\phi,T_s)= 0$ defines
the threshold of thermodynamic instability of uniform states, so
that for all states with temperatures below $T_s(\phi)$ we must
have that $\mathcal{E}_h(k;\phi,T) < 0$ not only at $k=0$, but
also at least within a finite interval $0 \le k \le k_0$. }

\begin{figure}
\begin{center}
\includegraphics[scale=.28]{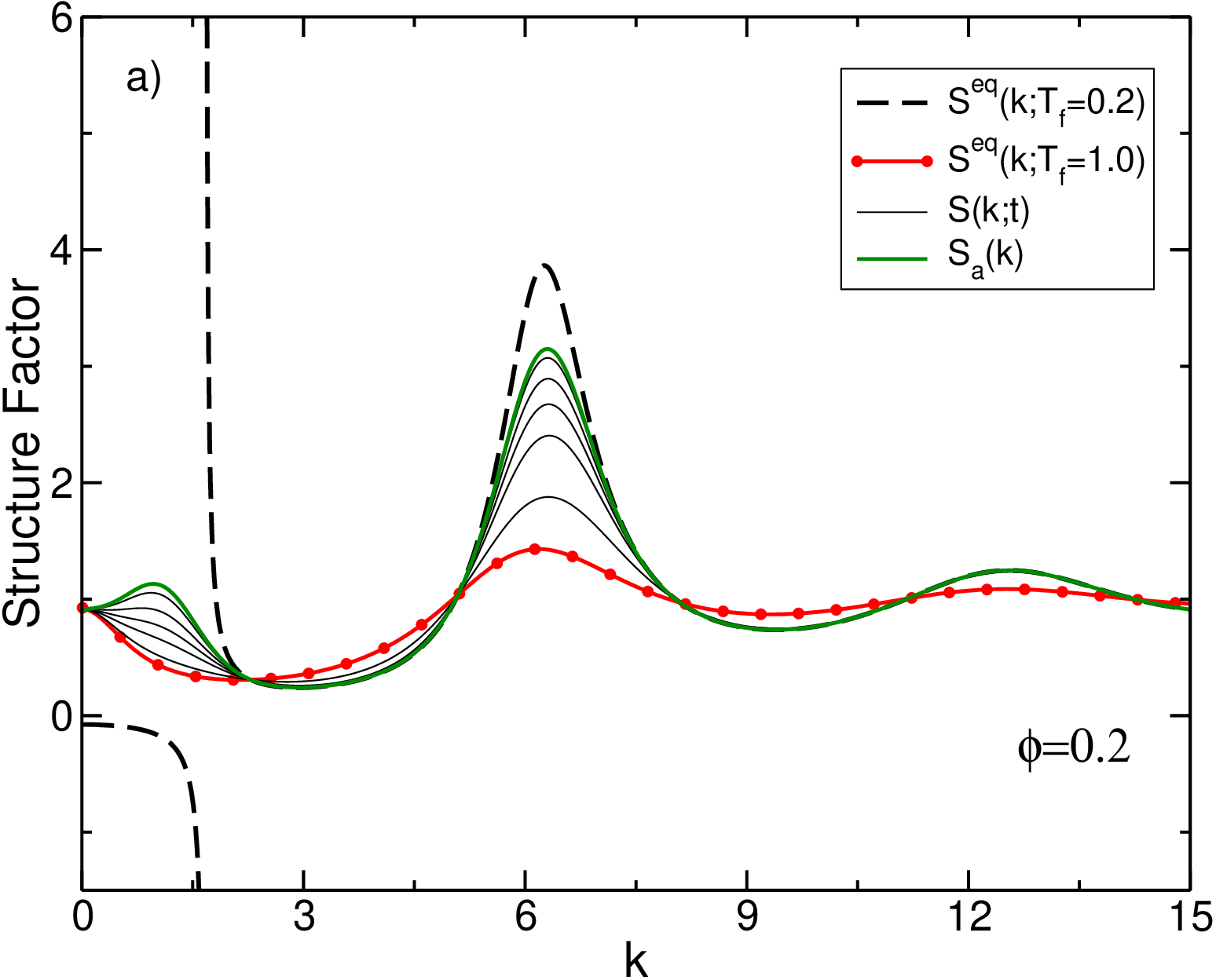}\hskip0.5cm
\includegraphics[scale=.28]{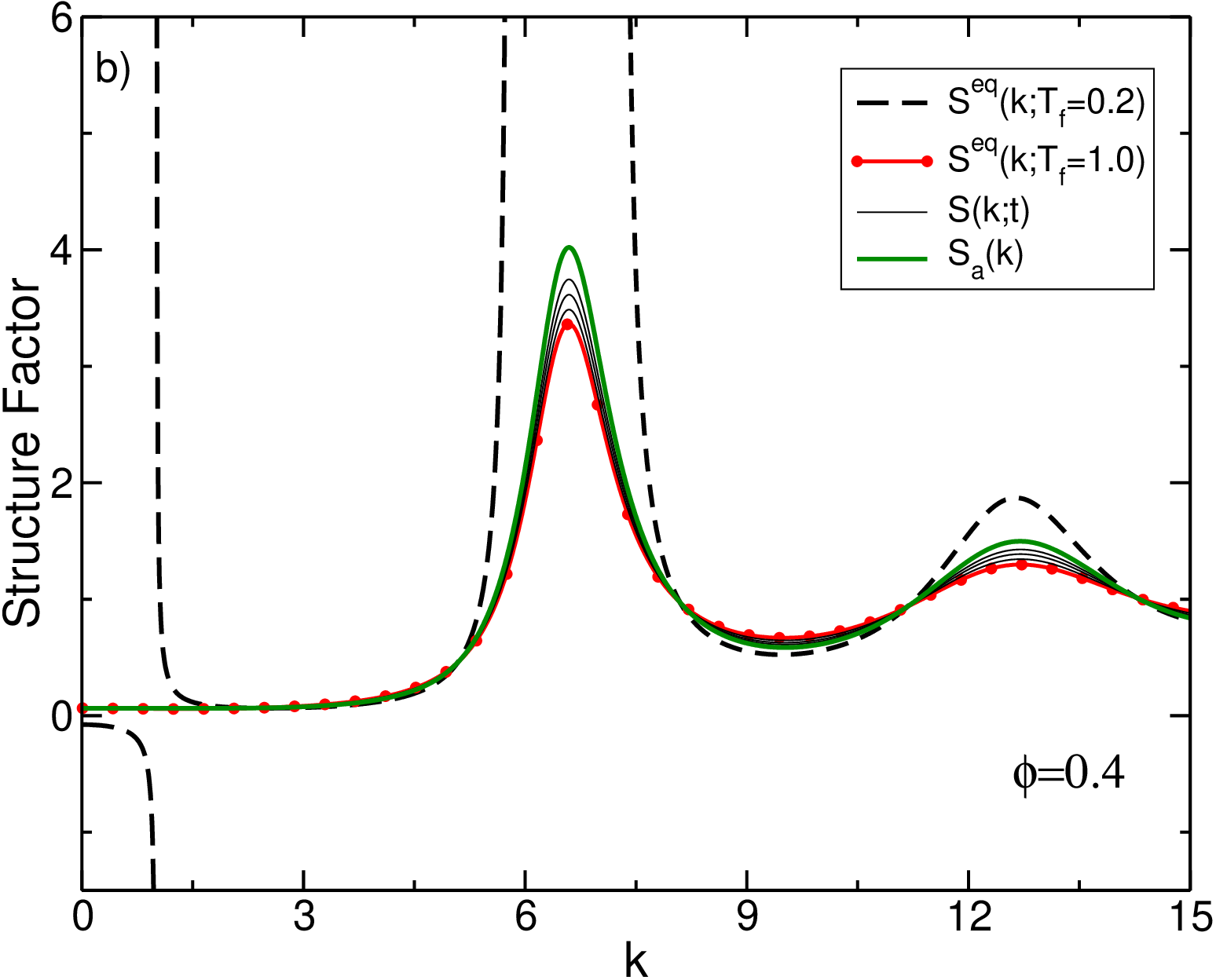}
\caption{\textcolor{black}{Initial equilibrium  static structure factor
$S^{(eq)}(k;\phi, T_i)= 1/\overline{n}\mathcal{E}_h(k;\phi, T_i)$
(dotted red curve) of the isochoric quench from the ergodic state
point $(\phi, T_i)$ to the state point $(\phi, T)$ inside the
spinodal region (panel (a) refers to the isochore  $\phi=0.2$ and
panel (b) to  $\phi=0.4$). The dashed line represents the
thermodynamic function $[\overline{n}\mathcal{E}_h(k;\phi,
T)]^{-1}$ at the latter point, whose negative and singular
features indicate the nonexistence of uniform equilibrium states
with static structure factor $S^{(eq)}(k;\phi, T)=
1/\overline{n}\mathcal{E}_h(k;\phi, T)$. The dark (green) solid
curves correspond to the (physically acceptable) asymptotic
long-time static structure factor $S_a(k) \equiv \lim_{t \to
\infty} S(k;t)=S^*(k;u_a)$ of this quench. The soft gray curves represent
$S(k;t)$ evaluated at an (arbitrary) sequence of values of the waiting time $t$.}} \label{fig7}
\end{center}
\end{figure}

\textcolor{black}{This means that the equilibrium static structure factor
$S^{(eq)}(k;\phi, T)= 1/\overline{n}\mathcal{E}_h(k;\phi, T)$ will
attain unphysical negative values in this interval and will
exhibit a singularity at $k= k_0$. This unphysical behavior, which
is a  manifestation of the non-existence of spatially uniform
equilibrium states inside the spinodal region, is illustrated in
Fig. \ref{fig7}, which plots $[\overline{n}\mathcal{E}_h(k;\phi,
T)]^{-1}$ (dashed curves) at the state points $(\phi,
T)=(0.2,0.2)$ and $(0.4,0.2)$ below the spinodal curve. One
might then expect that the non-equilibrium static structure factor
$S(k;t)$ and its asymptotic value $S_a(k)$, which are formally an exponential
interpolation between $S_i(k)$ and
$[\overline{n}\mathcal{E}_f(k;\phi,
T)]^{-1}$ (see Eqs.
(\ref{solsigmadkt}) and (\ref{solsigmadkta})), will inherit these singular
features of the function $[\overline{n}\mathcal{E}_f(k;\phi,
T)]^{-1}$. }

\textcolor{black}{In reality, this is not the case, since the appearance of
$[\overline{n}\mathcal{E}_f(k)]$ in the exponent of the
interpolating function, allows us to show, for example, that near
$k=k_0$  we have that $S(k;t)$ and $S_a(k)$  can be approximated, respectively, by
$S(k;t)=S^*(k;u(t))\approx S_i(k)+ 2k^2D_0 u(t)$ and $S_a(k) \approx S_i(k)+ 2k^2D_0 u_a$,
which are always positive definite. This is illustrated in Fig. \ref{fig7} with the results
for $S_a(k;\phi,T)$ (solid lines) for the quenches along the isochores $\phi=$ 0.2 and 0.4,
from the same initial temperature $T_i=1.0$, at which the initial static structure factor is
given by $S_i(k)=S^{(eq)}(k;\phi,T_i)=1/\overline{n}\mathcal{E}_h(k;\phi,T_i)$ (dotted lines),
to a final temperature $T=0.2$. As we can see in Figs.  \ref{fig7}(a) and (b), in both cases the
non-equilibrium stationary structure factor $S_a(k)$ does not exhibit any singular or
non-physical feature at any wave-vector.  Of course, for the same reason, the non-equilibrium
static structure factor $S(k;t)$ will evolve smoothly from $S_i(k)$ at $t=0$, to $S_a(k)$ as $t\to\infty$,
without exhibiting any hint of the singular behavior characteristic of the thermodynamic function
$[\overline{n}\mathcal{E}_f(k)]^{-1}$ below the spinodal curve. This is illustrated in Fig. \ref{fig7}
by the light-gray solid lines, corresponding to $S(k;t)$ evaluated at representative finite values
of the waiting time $t$. Notice in particular that for the quench illustrated in Fig. \ref{fig7}(a)
(corresponding to $\phi=0.2$), a mild but noticeable  small-$k$ peak of $S(k;t)$ develops at
$k_{max}(t)\ \approx 1$. This is to be contrasted with the quench along the isochore $\phi=0.4$
illustrated in Fig. \ref{fig7}(b), where this small-$k$ peak is imperceptible.}

\subsection{\textcolor{black}{Relationship with the theory of Cahn, Hilliard, and Cook.}}\label{subsectionIII.5}

\textcolor{black}{One of the main general features of
the non-equilibrium evolution of $S(k;t)$ after an isochoric
quench inside the spinodal region is precisely the development of this low-$k$
peak of $S(k;t)$ located at the time-dependent wave-vector $k_{max}(t)$. This peak is associated with the growing
length scale $\xi (t) \approx 2\pi/k_{max}(t)$ of the clusters
formed in the early stage of the process of spinodal decomposition. According to the
classical theory developed by Cahn, Hilliard, and Cook (CHC) \cite{cahnhilliard,cook}, and reviewed in detail by Furukawa  \cite{furukawa}, this small-$k$ peak evolves with waiting time $t$, moving to
progressively smaller wave-vectors, so that $\lim_{t\to\infty} \xi (t)=\infty$. The correct interpretation, however, is that this scenario only describes the very early stages of the phase separation process, up to a point in which the size $\xi (t)$ reaches mesoscopic dimensions so that additional effects (such as surface tension, convection, etc.), not contained in the theory, drive the system to full phase separation. }

\textcolor{black}{As it happens, the NE-SCGLE theory contains this classical CHC theory as its linear and small-$k$ limit. To see this more explicitly, let us notice that our fundamental time evolution equation for $S(k;t)$ in Eq.(\ref{relsigmadif2pp}) above, becomes Cook's equation (Eq. (3.4) in Ref. \cite{furukawa}) in the linear regime and the long-wavelength limit. The linear regime is achieved in our theory by neglecting the non-linear dependence of the time-dependent mobility $b(t)$ on $S(k;t)$ itself, i.e., by approximating $b(t)=1$, whereas in the long-wavelength limit (small $k$'s) we can expand the thermodynamic function $n\mathcal{E}_h(k;\phi,
T)$ in powers of $k$ up to quadratic order, $n\mathcal{E}_h(k;\phi,
T) \approx (a/k_BT)+(K/k_BT)k^2$, with $a=a(\phi,T)$ and $K=K(\phi,T)$ being wave-vector independent parameters. These two approximations,  with $M_0\equiv D^0/k_BT$, convert Eq. (\ref{relsigmadif2pp}) into}
\begin{equation}
\textcolor{black}{\frac{\partial S(k;t)}{\partial t} = 2k^2 M_0 \left[k_BT- (a+Kk^2)S(k;t)
\right],}  \label{relsigmadif2ppcook}
\end{equation}
\textcolor{black}{which is Eq. (3.4) of Ref. \cite{furukawa}. Notice also that Cook's approximate solution of this equation for shallow quenches (Eq. (3.9) of Ref. \cite{furukawa}) is identical to the approximation $S(k;t)=S^*(k;u(t))\approx S_i(k)+ 2k^2D_0 u(t)$ quoted above, provided that $b(t)=1$.}

\textcolor{black}{The fact that the NE-SCGLE theory contains the CHC theory as a particular limit also determines  the limitation of the  NE-SCGLE theory to describe the full phase separation process. Thus, for shallow quenches, where we know that full phase separation occurs, we understand that the  NE-SCGLE theory only describes the early stage of spinodal decomposition. Due to its non-linear nature, however, the NE-SCGLE theory complements this classical scenario (in which $\lim_{t\to\infty} \xi (t)=\infty$) with the prediction that $\lim_{t\to\infty} \xi (t)=\xi_a $, i.e., with the possibility that the process of spinodal decomposition is arrested when $\xi (t)$ reaches the length scale $\xi_a$ associated with the low-$k$ peak of the non-equilibrium asymptotic structure factor $S_a(k)$. Of course, if the predicted value of  $\xi_a$ is sufficiently large for  surface tension and convection to take over, the system will phase-separate completely. The possibility exists, however, that the predicted value of $\xi_a$  is not that large, or to be even of the scale of a few particle diameters. Under such circumstances the dynamic arrest of the phase separation process will occur right at its early stage, freezing the current structure and preventing the system from phase separating completely. These conditions are illustrated by the quench to $T=0.2$ along the isochore $\phi=0.2$, whose $S_a(k)$ is represented by the solid line of Fig. \ref{fig7}(a), for which $k_{max}\approx 1/\sigma$, and hence,  $\xi_a\approx 6\sigma$. Thus, we need a criterion to locate the crossover temperature $T_0(\phi)$ below which the predicted dynamic arrest will in fact occur, and above which full phase separation will be driven by effects not contained in our equations. In the rest of this section we shall search for such criterion in the temperature dependence of the asymptotic structure represented by $S_a(k)$.}

\subsection{\textcolor{black}{Non-equilibrium small-$k$ peak of $S_a(k;\phi, T)$ and diverging length scale $\xi_a(\phi, T)$. }}\label{subsectionIII.5}

\begin{figure}
\begin{center}
\includegraphics[scale=.28]{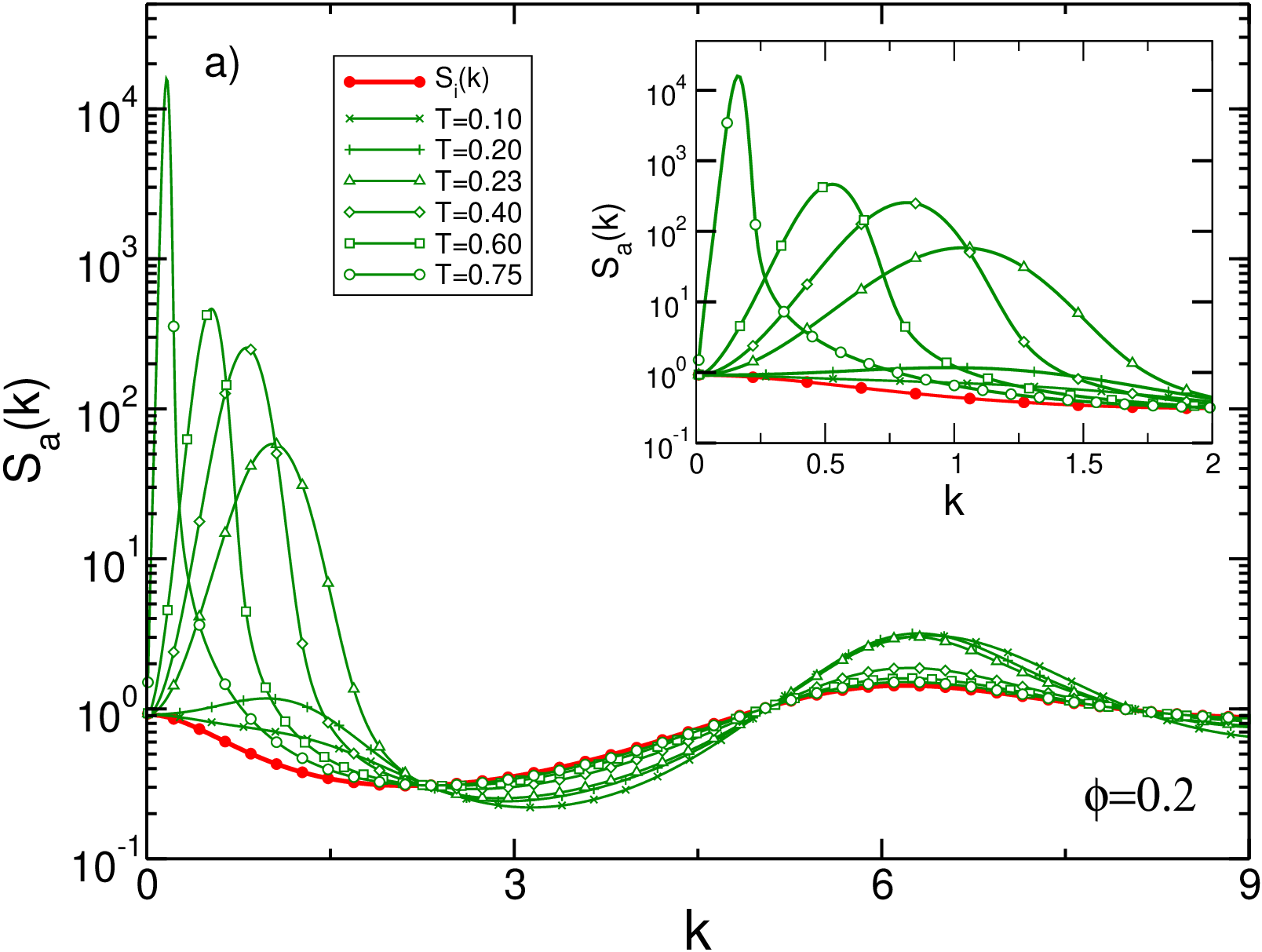}\hskip0.5cm
\includegraphics[scale=.28]{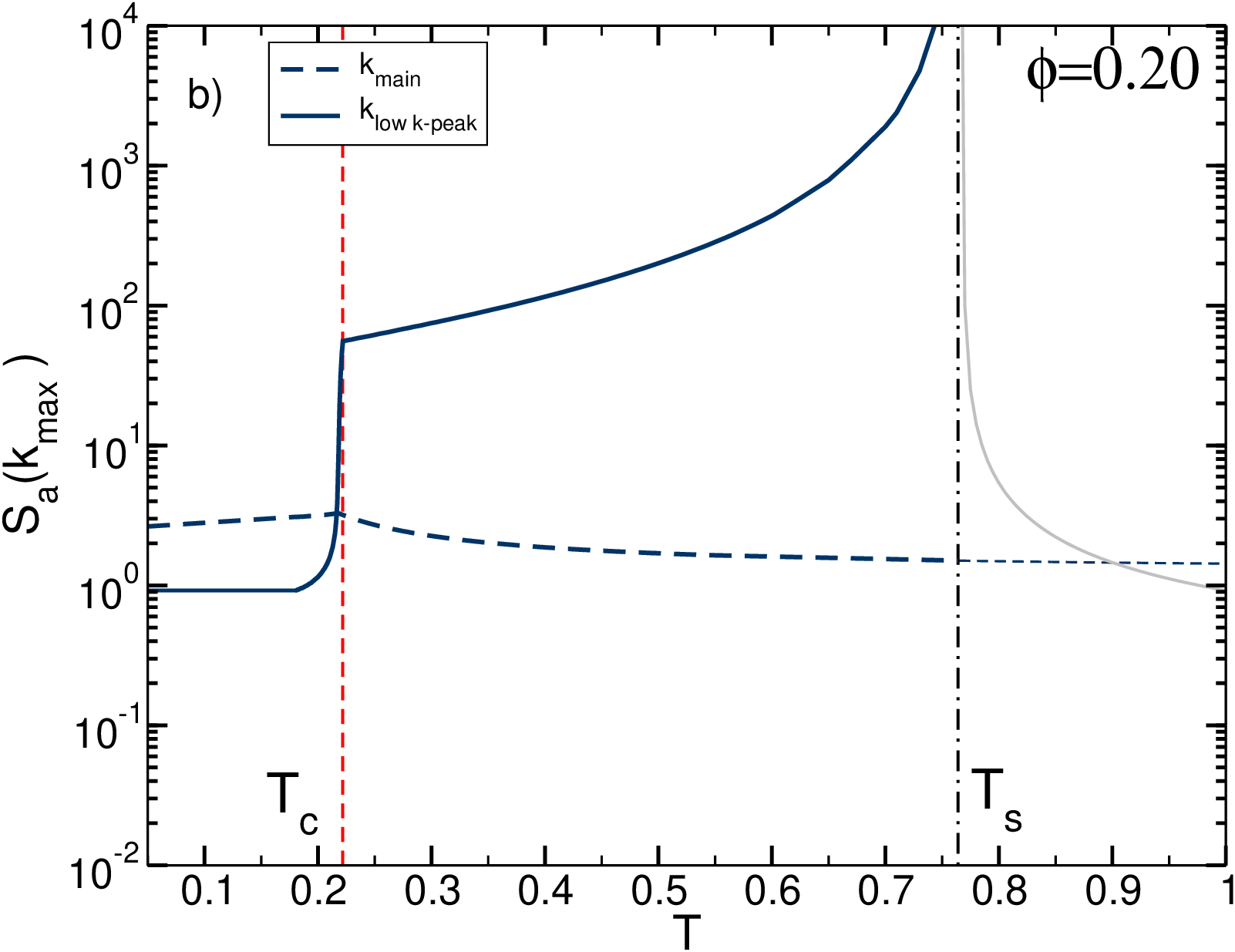}
\caption{\textcolor{black}{(a) Non-equilibrium
stationary structure factor $S_a(k;\phi, T)$ as a function of
wave-vector $k$ for a sequence of instantaneous isochoric cooling processes   along the isochore $\phi=0.2$ from the initial temperature $T_i=1.0$ to the indicated final temperature T. The figure illustrates the $T$-dependence of the low-$k$ peak ($k\approx 1/\sigma$; also zoomed in the inset) and of the main peak ($k\approx 2 \pi/\sigma$) of  $S_a(k;\phi, T)$. (b) Height of the low-$k$ peak (dark solid line) and of the main peak (dark dashed line) of $S_a(k;\phi, T)$, as a function of the final temperature $T$  along the isochore $\phi=0.2$. The soft gray solid line represents the  diverging behavior (when $T \to T_s$ from above) of the $k=0$ value of the equilibrium static structure factor, $S^{(eq)}(k=0;\phi, T)$.}} \label{fig20}
\end{center}
\end{figure}

\textcolor{black}{Let us thus analyze in more detail the dependence of  $S_a(k;\phi, T)$ on the depth of the quench, i.e., on the final temperature $T$, as well as  on the volume fraction $\phi$. With this intention, in Fig. \ref{fig20}(a) we present a set of results for $S_a(k;\phi, T)$ that illustrate the $T$-dependence of this asymptotic structural property  for quenches along the isochore $\phi=$ 0.2. All of these quenches start with the system equilibrated at the same initial temperature $T_i=1.0$, whose corresponding equilibrium static structure factor $S_i(k)= S^{(eq)}(k;\phi=0.2, T=1.0)$ is represented by the dotted curve. Each quench ends at a different final temperature $T$, and the solid lines represent the resulting $S_a(k;\phi=0.2, T)$. Thus, the dotted line and the solid line corresponding to $S_a(k;\phi=0.2, T=0.2)$  are the same as the dotted and solid lines of Fig. \ref{fig7}(a), but now in the vertical axis we use logarithmic scale to visualize the strong increase in $S_a(k_{max};\phi, T)$  as the final temperature $T$ approaches the spinodal temperature $T_s(\phi=0.2)=0.76$ from below. This is clearly the most visible trend exhibited by this set of curves, together with the fact that the position $k_{max}(\phi, T)$ of this growing non-equilibrium peak of $S_a(k;\phi, T)$ moves monotonically to the left, so that the corresponding length scale $\xi_a(\phi, T)=2\pi/k_{max}(\phi, T)$ increases with increasing $T$. }

\textcolor{black}{In fact, to further visualize these trends, in Fig. \ref{fig20}(b) we plot as a solid line the height $S_a(k_{max};\phi, T)$  of the low-$k$ peak of $S_a(k;\phi, T)$ as a function of the final temperature $T$ along the isochore $\phi=0.2$. This curve clearly illustrates that the behavior of $S_a(k_{max};\phi, T)$ is distinctly different in the regime where the final temperature falls below the glass-glass transition temperature $T_c(\phi=0.2) = 0.22$, and in the regime where $T$ falls in the interval $T_c \le T \le T_s$. In the former, for the lowest temperatures $S_a(k;\phi, T)$ starts almost constant, with a value similar to $S_i(k=0)$ (recall that  $S_a(k_{max}) \approx S_i(k_{max})+ 2k_{max}^2D_0 u_a\approx S_i(k_{max}) \approx S_i(0) \approx 1$), but as $T$ approaches  $T_c$ from below it  increases sharply to a value $S_a(k;\phi=0.2, T_c^-)=55$. In the interval $T_c \le T \le T_s$, on the other hand, $S_a(k_{max})$ starts from the value $S_a(k;\phi=0.2, T_c^+)=55$ (i.e., $S_a(k;\phi, T)$ is continuous at $T=T_c$), and increases monotonically up to the spinodal temperature $T_s$, at which it diverges as $(T_s-T)^{-\mu}$, with $\mu \approx 1.75$. This divergence of $S_a(k_{max};\phi, T)$  is the non-equilibrium counterpart of the divergence of $S^{(eq)}(k=0;\phi, T)=[\partial(p/k_BT)/\partial n]_T^{-1}$ when $T$ approaches $T_s$ from above. For completeness, although it has no direct relevance regarding the possibility of dynamic arrest, the divergence of $S^{(eq)}(k=0;\phi, T)$ is also shown in Fig. \ref{fig20}(b) (soft gray solid curve to the right of $T_s$). }

\begin{figure}
\begin{center}
\includegraphics[scale=.18]{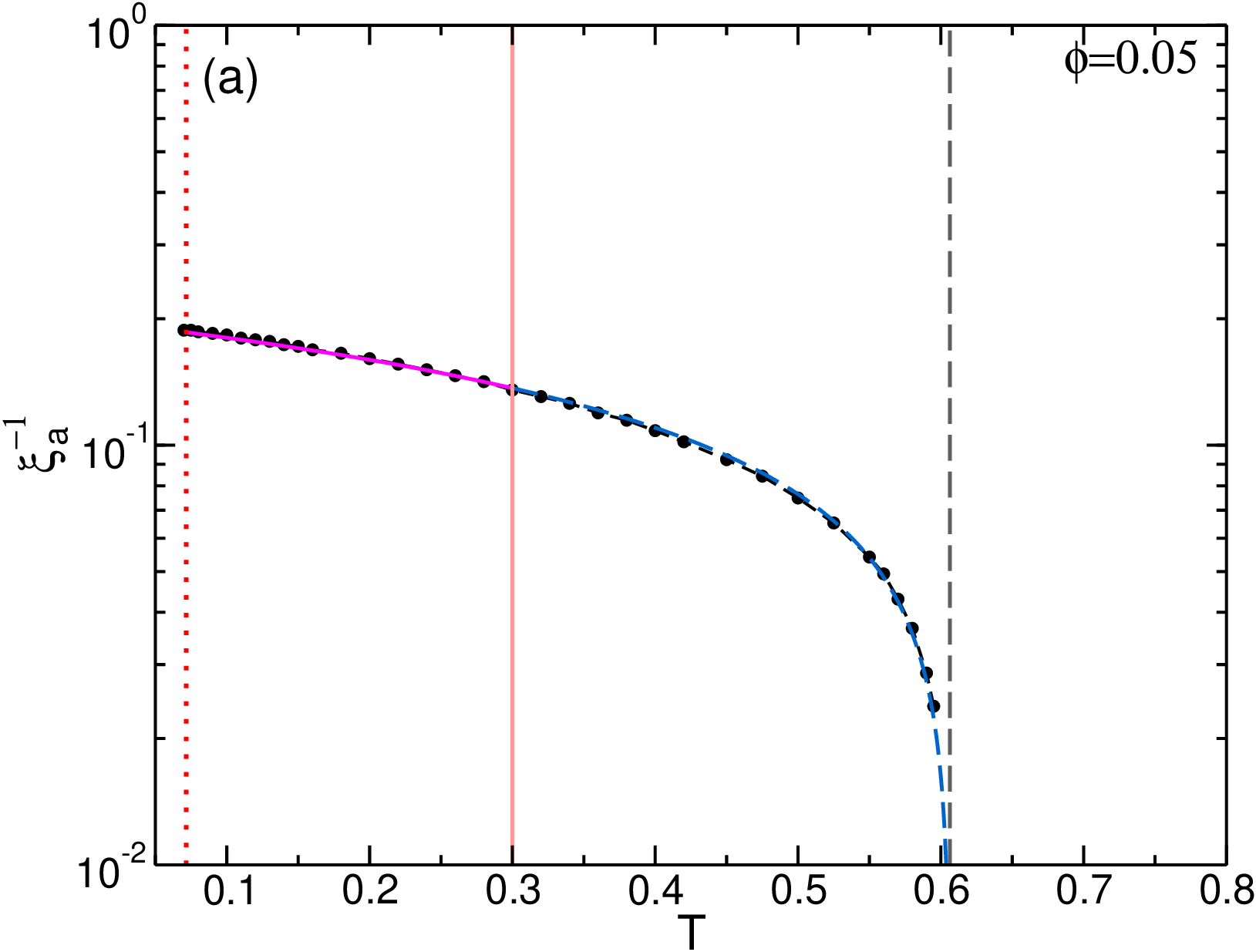}\hskip.2cm
\includegraphics[scale=.18]{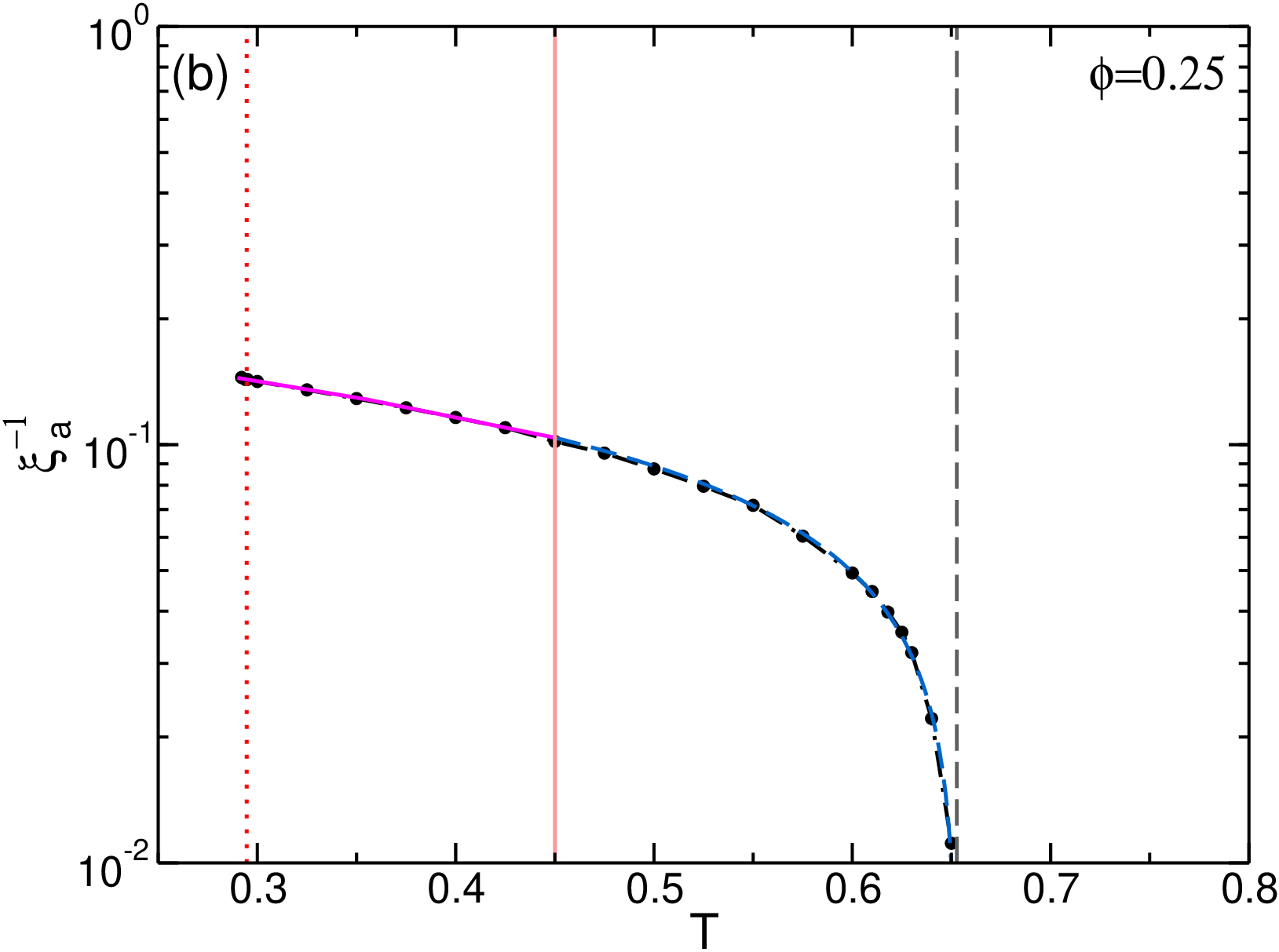}\hskip.2cm
\includegraphics[scale=.18]{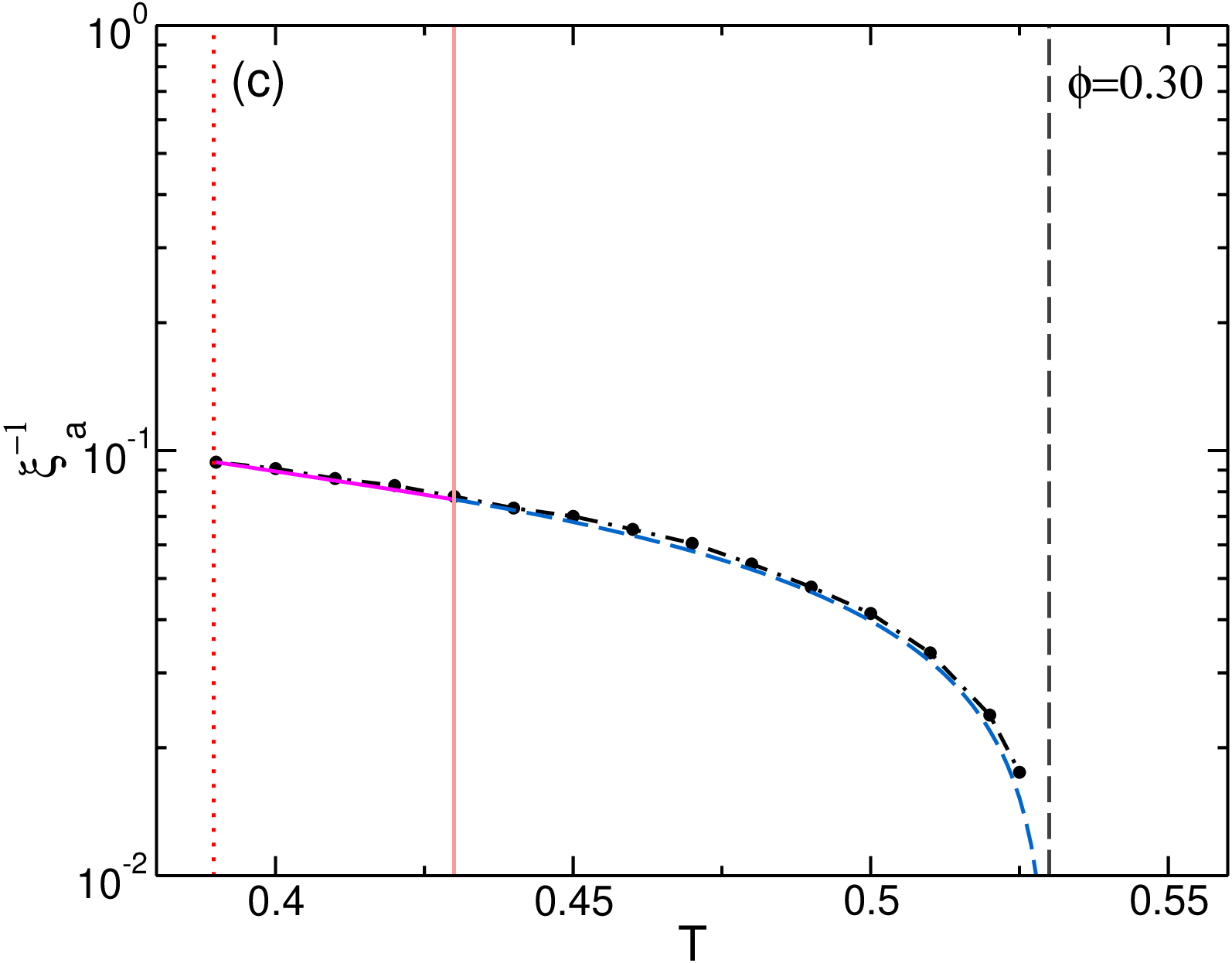}\hskip.2cm
\caption{\textcolor{black}{Characteristic length $\xi_a(\phi, T)=2\pi/k_{max}(\phi,
T)$ associated with the low-$k$ peak of $S_a(k_{max};\phi, T)$, as a function
of the final temperature T of the quench,  for the isochores (a)
$\phi=0.05$,  (b) $\phi=0.25$, and (c) $\phi=0.30$. The dots represent $\xi_a(\phi, T)$ determined directly from  $S_a(k_{max};\phi, T)$ at a discrete set of temperatures, while the composed (solid-dashed) curve is the fit of these numerical data obtained with the parametrization expression in Eq. (\ref{fitcsi}). The vertical lines represent the temperatures $T_c(\phi) < T_0(\phi) <T_s(\phi)$, with $T_0(\phi)$ indicating the crossover of the $T$-dependence of $\xi_a(\phi, T)$ from exponential (solid magenta portion) to diverging power-law (dashed blue portion).} }
\label{fig9}
\end{center}
\end{figure}

\textcolor{black}{Let us also notice in Fig. \ref{fig20}(a) that in the regime $T_c \le T \le T_s$, while $S_a(k_{max})$ increases  and diverges as $T \to T_s^-$, its position $k_{max}(\phi, T)$ moves to the left and vanishes at $T=T_s$, so that the corresponding length scale $\xi_a(\phi,T)\equiv 2\pi/k_{max}(\phi,T)$ also diverges as  $T$ approaches $T_s$ from below. This divergence is the non-equilibrium counterpart of the divergence of the equilibrium correlation length $\xi^{eq}(\phi,T)$ as $T \to T_s^+$, predicted in 1914 by Ornstein and Zernike \cite{oz1914} to explain the phenomenon of critical opalescence along the critical isochore. In Fig. \ref{fig9} we present the temperature dependence of $\xi_a(\phi, T)$  in the interval $T_c \le T \le T_s$ along three illustrative isochores. We notice that in this interval  $\xi_a(\phi, T)$ exhibits two sub-regimes, characterized by a different dependence  on the final temperature. In the first sub-regime $\xi_a(\phi, T)$ increases exponentially with $T$, whereas in the second, which we shall
refer to as the ``Ornstein-Zernike domain'',  $\xi_a(\phi, T)$
increases with $T$ as an inverse power of the reduced temperature
$T^*\equiv (T_s-T)/T_s$, diverging at the spinodal line with an
exponent close to the classical (Ornstein-Zernike) exponent that
describes the critical divergence of the equilibrium correlation
length \cite{oz1914}. }

\textcolor{black}{ In fact, we have noticed that the
results for $\xi_a(\phi, T)$ can be conveniently parametrized with an empiric
expression that matches an exponential function below $T_0$ with a power law above $T_0$, in a manner that both, the function and its derivative, are continuous at $T_0$. Such an expression reads
\begin{align}
\xi_a(T)&  =\xi_a(T_c^+)e^{\nu
\left(\frac{T-T_c}{T_s-T_0}\right)},\text{ \ \ \ \ \ \ \ \ \ \ \ \
\ \ \ \ \ \ \ \ for \ \
 }\,T_c\le T\le T_0
\nonumber\\
&  =\xi_a(T_c^+)e^{\nu
\left(\frac{T_0-T_c}{T_s-T_0}\right)}\left[\frac{T_s-T}{T_s-T_0}\right]^{-\nu},
\text{ \ \ for \ \ }\,T_0\le T\le T_s.
\label{fitcsi}%
\end{align}
Since we already have the theoretical prediction for the spinodal and the glass-glass transition lines $T_s(\phi)$ and  $T_c(\phi)$, and we can determine the value $\xi_a(T_c^+)$ of  $\xi_a(T)$ immediately above $T_c$, this functional form allows us to determine the exponent  $\nu$ and the crossover temperature $T_0(\phi)$ by the best fit of the numerical data of $\xi_a(T)$. In fact, we actually fixed the exponent $\nu$ to its classical OZ value  $\nu=0.5$, so that the only parameter to be determined by this fit is the crossover temperature $T_0(\phi)$. As observed in Fig.  \ref{fig9}, the quality of this parametrization of the numerical results for $\xi_a(\phi,T)$ along the three illustrative isochores is excellent.}

\subsection{\textcolor{black}{The threshold temperature $T_0(\phi)$ and the gelation line.}}\label{subsectionIII.5}

\textcolor{black}{The procedure just described, performed at a sequence of isochores, defines a  curve $T_0(\phi)$ in the dynamic arrest diagram, which divides the interval $T_c(\phi)\le T\le T_s(\phi)$ in  two well-defined sub-regimes: the subinterval $T_c(\phi)\le T\le T_0(\phi)$, where $\xi_a(T)$ increases exponentially with $T$,  and the subinterval  $T_0(\phi)\le T \le T_s(\phi)$, where $\xi_a(T)$ enters its diverging Ornstein-Zernike regime. Although we cannot jump immediately to the conclusion that these two subintervals are to be identified, respectively, with the gel and the full spinodal decomposition regions of state space, these findings provide important elements to consider in building a sound physical interpretation of the dynamic arrest diagram in Fig. \ref{fig3b}. The expectation, however, is that indeed this crossover curve will eventually be identified with the gelation line.}

\begin{figure}
\begin{center}
\includegraphics[scale=.18]{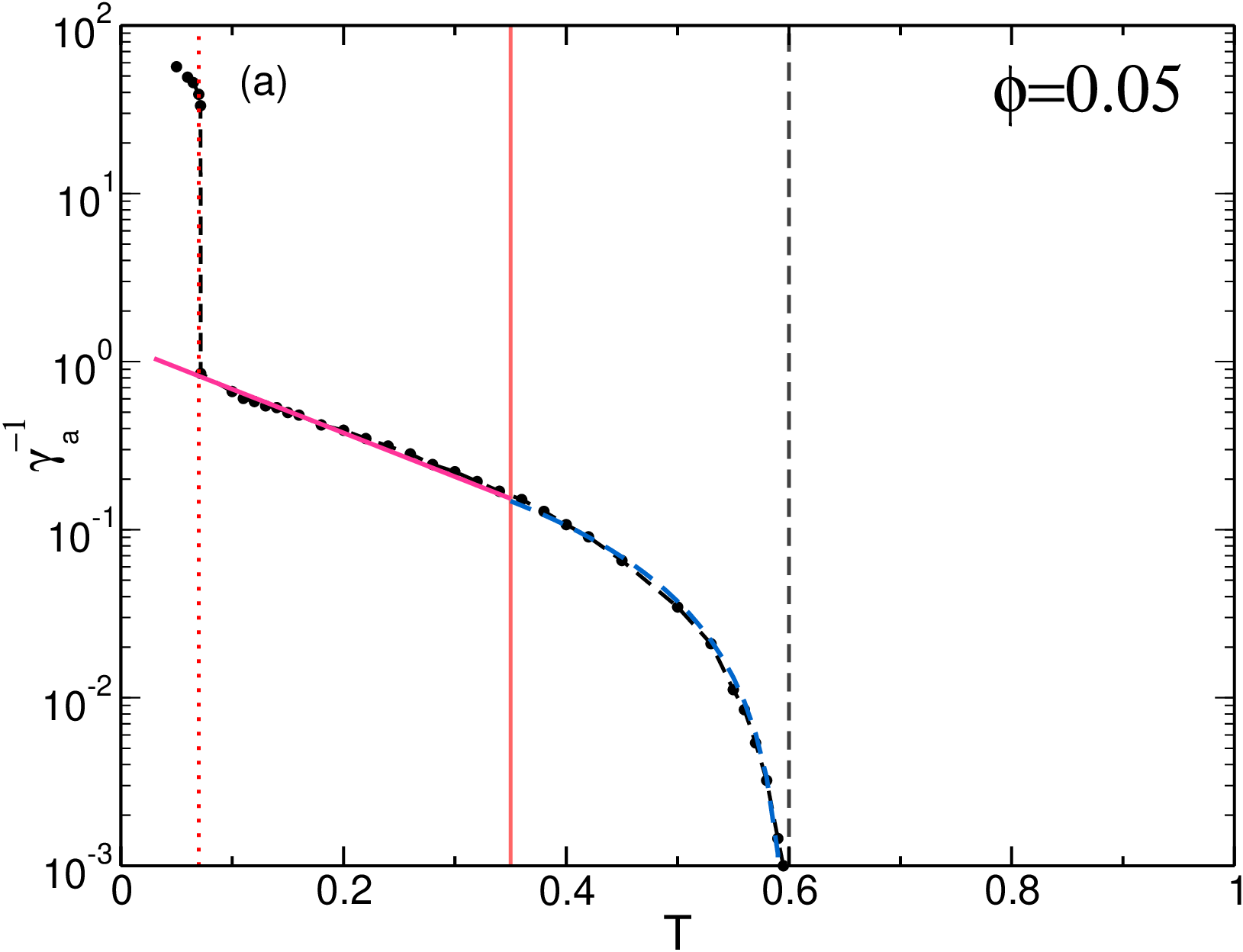}\hskip.2cm
\includegraphics[scale=.18]{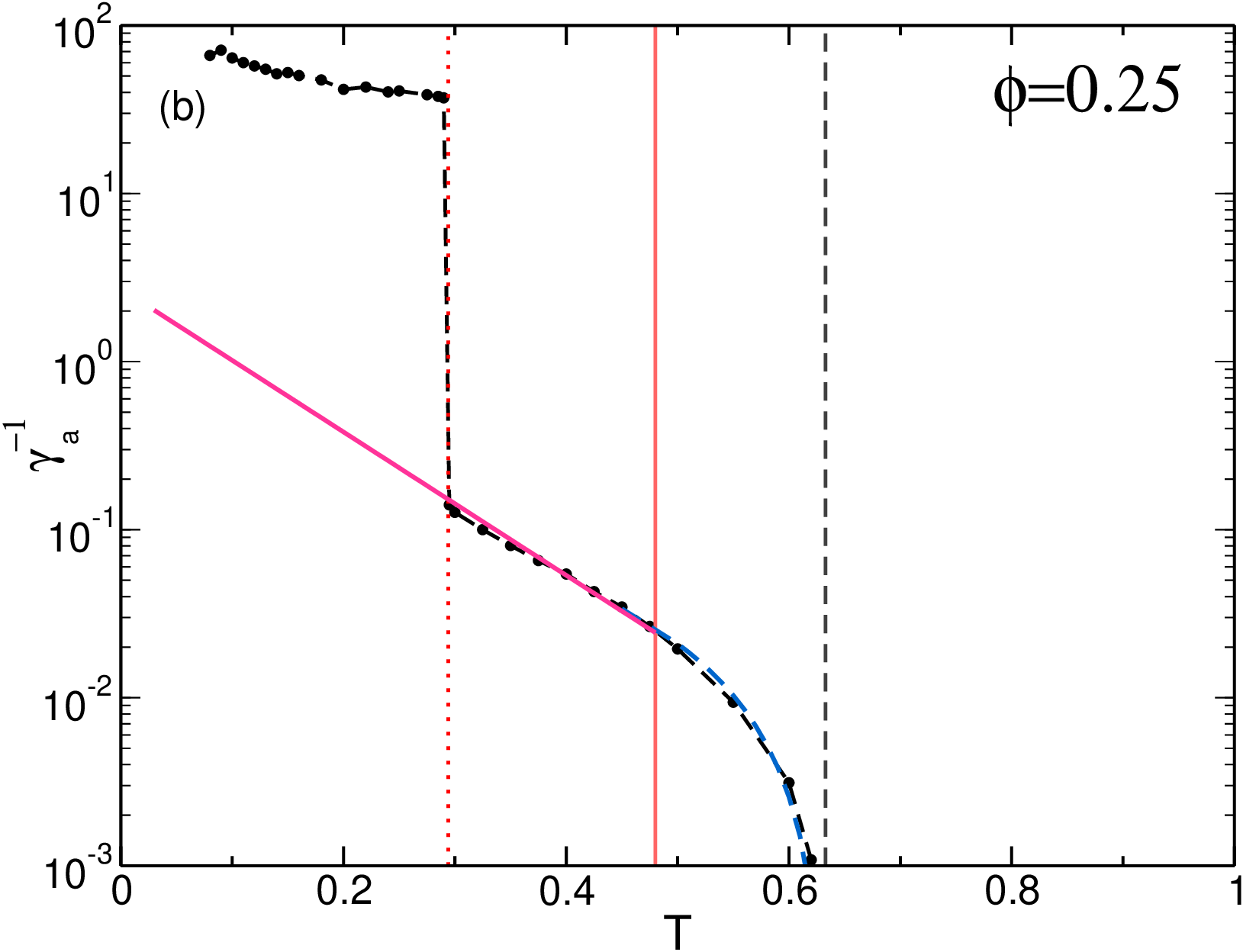}\hskip.2cm
\includegraphics[scale=.18]{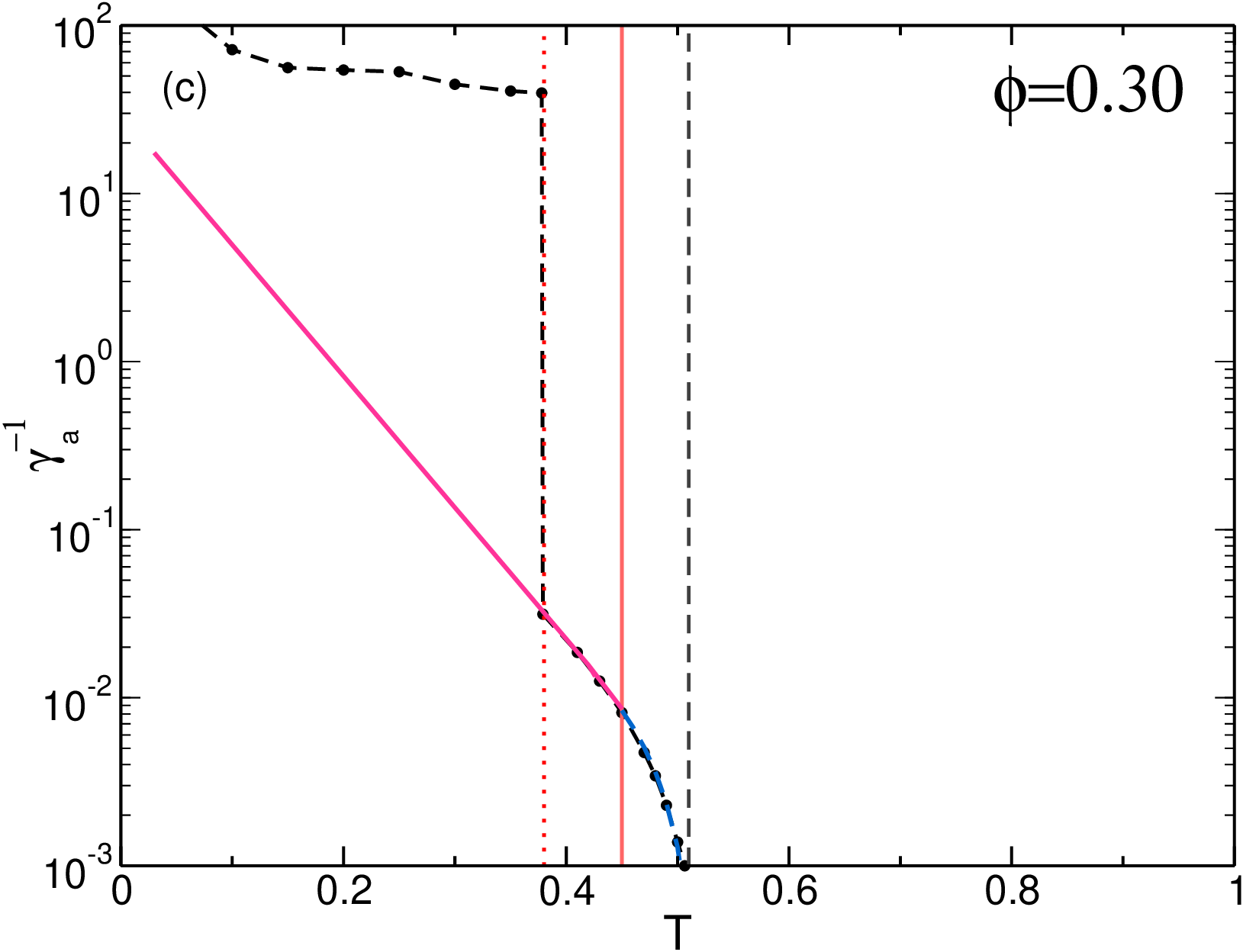}\hskip.2cm
\caption{Dynamic order  parameter $\gamma^{*}_a (\phi,T)$ as a
function of the final temperature $T$, and its parametrization
according to Eq. (\ref{fitgamma}) at the isochores (a)
$\phi=0.05$,  (b) $\phi=0.25$, and (c) $\phi=0.30$. \textcolor{black}{ As in Fig. \ref{fig9}, the vertical lines represent the temperatures $T_c(\phi) < T_0(\phi) <T_s(\phi)$, with $T_0(\phi)$ indicating the crossover of the $T$-dependence of $\gamma_a(\phi, T)$ from exponential (solid magenta portion) to diverging power-law (dashed blue portion)}}
\label{fig5}
\end{center}
\end{figure}

\textcolor{black}{Further elements supporting the notion of the existence of this crossover are also revealed by the behavior of other properties. In fact, going back to Fig. \ref{fig3}, one can immediately appreciate that at least along the isochore   $\phi=0.2$, also the asymptotic square localization length $\gamma^*_a(T)$ seems to exhibit a similar behavior as $\xi_a(T)$. To test this idea, let us propose a parametrization of its $T$-dependence  similar to that employed for $\xi_a(T)$, namely,}
\begin{align}
\gamma^*_a(T)&  =\gamma^{*}_a (T_c)e^{\alpha
\left(\frac{T-T_c}{T_s-T_0}\right)},\text{ \ \ \ \ \ \ \ \ \ \ \ \
\ \ \ \ \ \ \ \ for \ \
 }\,T_c\le T\le T_0
\nonumber\\
&  =\gamma^{*}_a (T_c)e^{\alpha
\left(\frac{T_0-T_c}{T_s-T_0}\right)}\left[\frac{T_s-T}{T_s-T_0}\right]^{-\alpha},
\text{ \ \ for \ \ }\,T_0\le T\le T_s.
\label{fitgamma}%
\end{align}
We found that a reasonable representation of the
numerical results for $\gamma^*_a(T)$ in Fig. \ref{fig3} could
be obtained with the exponent $\alpha$ fixed as $\alpha=1.5$, and
adjusting $T_0$ to find the best overall fit, thus leading to the
value $T_0(\phi=0.2)=0.52$.  We tested the same procedure for other isochores,
also fixing the exponent $\alpha$ as $\alpha=1.5$, and determined
$T_0(\phi)$ from the best overall fit. Fig. \ref{fig5} illustrates
the accuracy of this fit at other three illustrative isochores.
\textcolor{black}{This procedure provides a second determination of the location of the crossover curve $T_0(\phi)$.}

\textcolor{black}{ Let us finally mention that not only the spatially-meaningful length scales $\xi_a(\phi,T)$ and $\sqrt{\gamma^{*}_a (\phi,T)}$ exhibit this dual temperature dependence. Looking back again at Fig. \ref{fig3}, we see that the behavior of the ``material time'' $u_a(\phi,T)$ also suggests a similar pattern. Although we do not provide further details, let us mention that we have carried out the same parametrization exercise, with a similar composed functional form, leading to similar accuracy of the overall fits. Although in this case the power-law exponent does depend slightly on volume fraction, this parametrization with the exponent at the value of 2, does provide still a third route to determine the crossover temperature $T_0(\phi)$. Of course, there is no reason to expect full quantitative consistency between these three manners to determine  $T_0(\phi)$. Nevertheless, they coincide in the general notion that the region bounded by the spinodal curve $T_s(\phi)$ from above and by the glass-glass transition line $T_s(\phi)$ from below, is divided in two physically distinct domains. }

\subsection{\textcolor{black}{High-density ``ordinary'' liquid-glass transition.}}\label{subsectionIII.5}

\textcolor{black}{Let us also notice from Fig. \ref{fig20}(a) that besides the development of this low-$k$ spinodal-decomposition peak,  the asymptotic structure $S_a(k;\phi,T)$ also exhibits its main meak at $k\approx 2 \pi/\sigma$, associated with the short-range  nearest-neighbor arrangement among the particles. This main peak also varies with the final temperature $T$, but in a far less dramatic manner. Thus, we see that the height of the main peak, plotted as the dashed line of  Fig. \ref{fig20}(b), increases only slightly as the final temperature increases from 0 to $T_c$, and then decreases monotonically in the interval $T_c \le T \le T_s$. This behavior has some form of continuation in the equilibrium regime (above the spinodal temperature)  in the height of the main peak of the equilibrium static structure factor $S^{(eq)}(k;\phi, T)$ (soft gray solid line for $T \ge T_s$). Thus, if the importance of these two competing modes is measured by the height of the respective peak of $S_a(k;\phi,T)$, then we conclude that for shallow quenches inside the spinodal region, but at volume fractions below $\phi_b$, the asymptotic structure is dramatically dominated by the small-$k$ peak, which diverges as we approach the spinodal line from below.  Along the same isochores, however, deeper quenches (below the glass-glass transition temperature $T_c$) lead to an opposite scenario, in which the asymptotic structure is dominated by the main peak of $S_a(k;\phi,T)$.  With the aim of understanding the nature of this glass-glass transition, let us first discuss its relationship with the ``ordinary'' liquid-glass transition predicted to occur at $T_c(\phi)$, but for volume fractions above the bifurcation point, $\phi\ge \phi_b$.}

\begin{figure}
\begin{center}
\includegraphics[scale=.28]{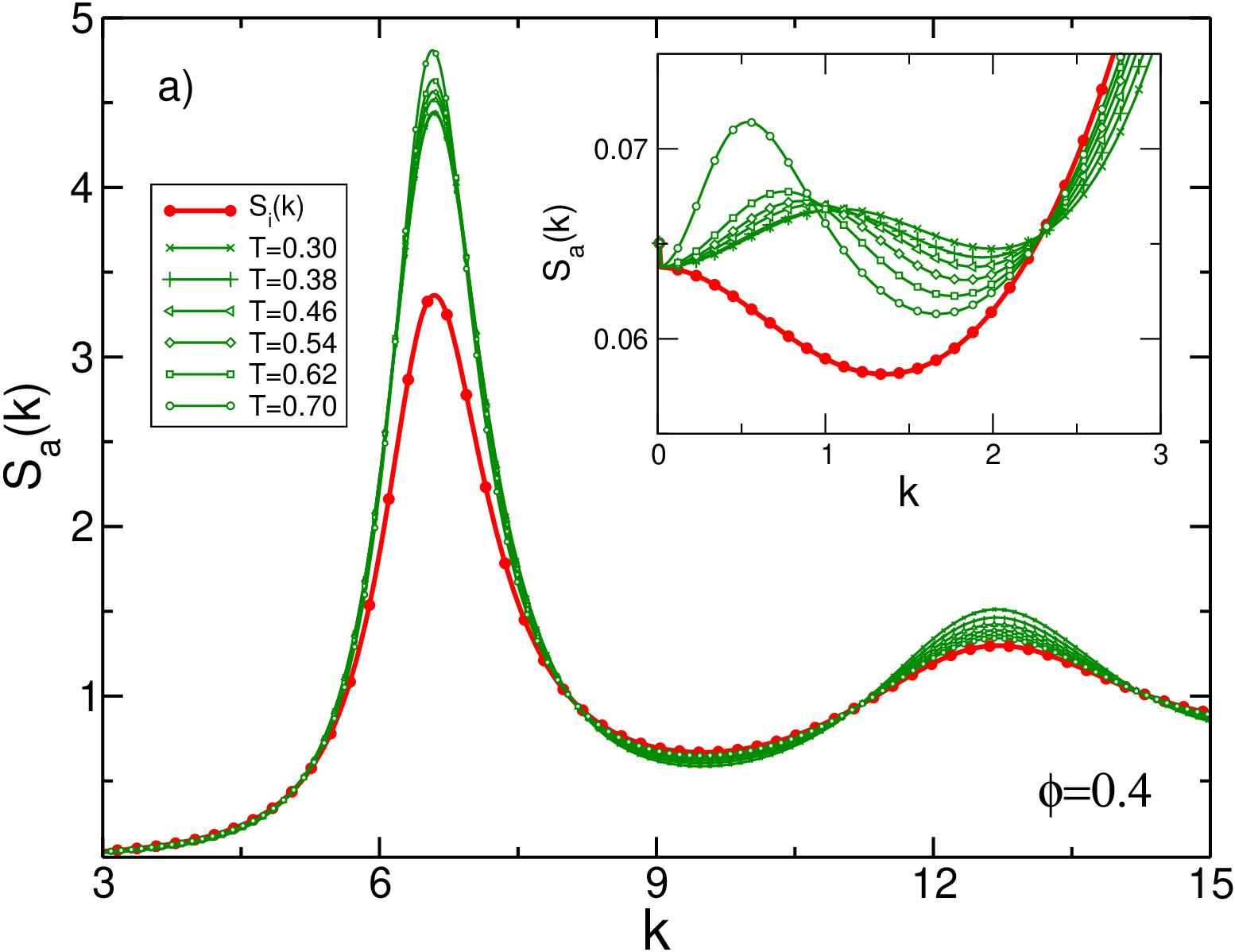}\hskip0.5cm
\includegraphics[scale=.28]{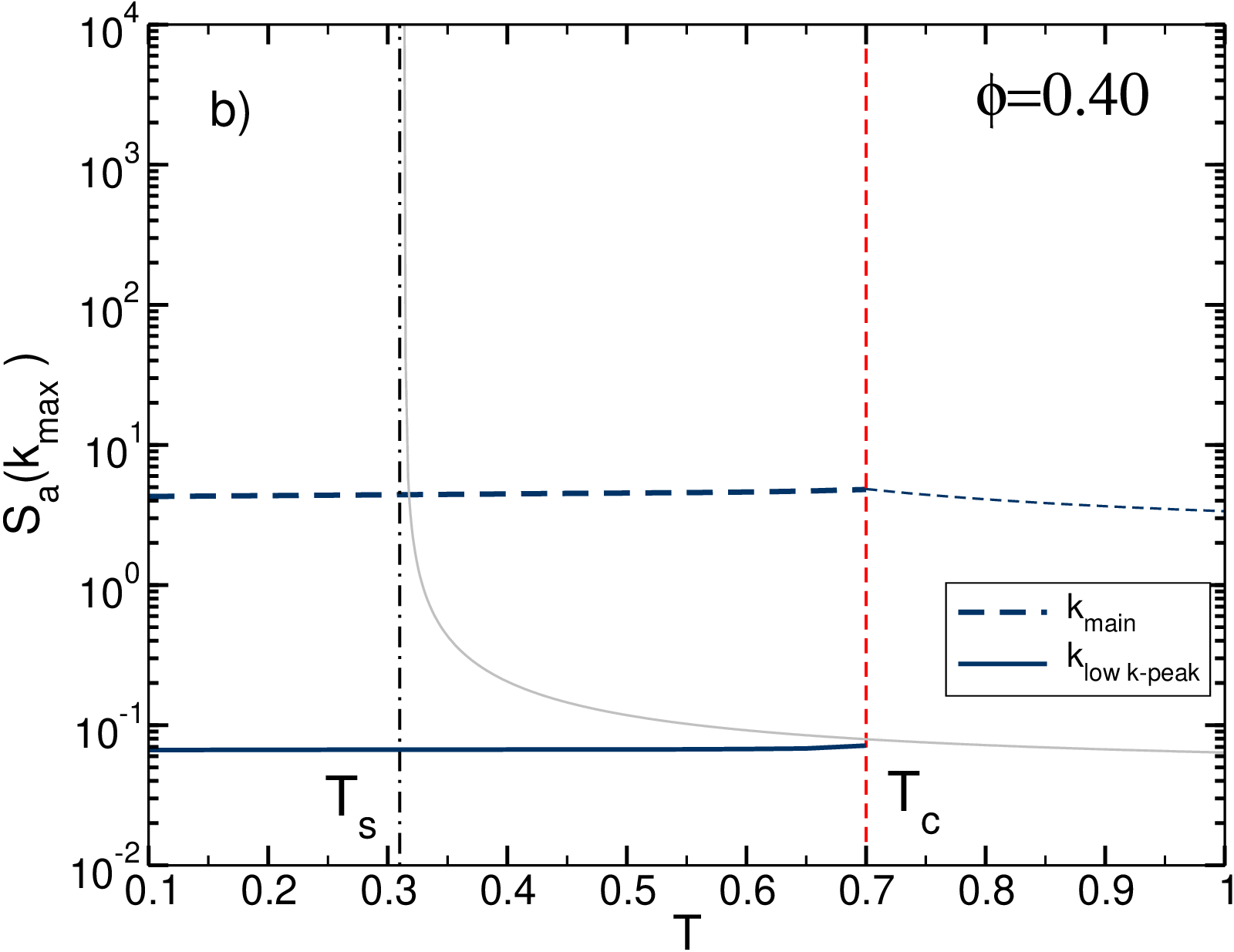}
\caption{\textcolor{black}{(a) Non-equilibrium
stationary structure factor $S_a(k;\phi, T)$ as a function of
wave-vector $k$ for a sequence of instantaneous isochoric cooling processes   along the isochore $\phi=0.4$ from the initial temperature $T_i=1.0$ to the indicated final temperature T. The figure illustrates the $T$-dependence of the low-$k$ peak ($k\approx 1/\sigma$; also zoomed in the inset) and of the main peak ($k\approx 2 \pi/\sigma$) of  $S_a(k;\phi, T)$. (b) Height of the low-$k$ peak (dark solid line) and of the main peak (dark dashed line) of $S_a(k;\phi, T)$, as a function of the final temperature $T$  along the isochore $\phi=0.4$. The soft gray solid line represents the  diverging behavior (when $T \to T_s$ from above) of the $k=0$ value of the equilibrium static structure factor, $S^{(eq)}(k=0;\phi, T)$. }} \label{fig21}
\end{center}
\end{figure}

\textcolor{black}{Thus, Fig. \ref{fig21} presents similar results as Fig. \ref{fig20}, but now at the isochore  $\phi=$ 0.4, representative of volume fractions well above the volume fraction  $\phi_b=$ 0.33 of the bifurcation point. Clearly, the scenario is now completely different from the previous one. As we can see in Fig. \ref{fig21}(a), the main difference is that now the small-$k$ peak is virtually absent. In reality this peak is still there, but it is imperceptible in the scale of the main figure, as revealed by the rather extreme zoom in the inset. Thus, in the virtual absence of this spinodal peak, the most visible feature of $S_a(k;\phi,T)$ is its main peak located at $k\sigma \approx 2\pi$. As illustrated by the set of plots  of $S_a(k;\phi,T)$ in Fig. \ref{fig21}(a), the height of this main peak exhibits a rather mild variation as a function of the final temperature $T$. This is also illustrated by the dark blue dashed line of Fig. \ref{fig21}(b). }

\textcolor{black}{ Let us notice that along the isochores with volume fraction $\phi < \phi_b$, the dominating feature is the existence of the  interval  $T_c \le T \le T_s$, where the spinodal decomposition peak determines the nature of the process of dynamic arrest. This interval naturally defines the other two complementary regions of state space: the high-temperature equilibrium domain above the spinodal curve, and the low-temperature glass region below $T_c(\phi)$. With this scenario in mind, it is natural to question if at isochores with volume fraction above $\phi_b$, where the order $T_c < T_s$ has been inverted to $T_s < T_c$, we will again find three physically distinct temperature regimes: the high-temperature  equilibrium domain above $T_c(\phi)$, the window $T_s \le T \le T_c$, and the low-temperature domain below the spinodal curve $T_s(\phi)$. In reality, the NE-SCGLE answer to this question was already given by the behavior of $\gamma_a$ in Fig. \ref{fig2}, which only identifies two physically distinct regions, namely, the  high-temperature  equilibrium liquid domain, $T>T_c(\phi)$, and the region of dynamically arrested states, $0 \le T \le T_c$, without the spinodal temperature $T_s(\phi)$ having any relevance. }

\textcolor{black}{ In an attempt to better understand this scenario, let us notice that along the isochores with volume fraction larger than $\phi_b$, we have that the equilibrium static structure factor $S^{(eq)}(k;\phi, T)$ exists in the interval $T \ge T_s(\phi)$ ($S^{(eq)}(k=0;\phi, T)$ is the soft gray solid line in Fig. \ref{fig21}(b)), and the dynamically-arrested static structure factor $S_a(k;\phi,T)$ exists in the interval $T\le T_c(\phi)$ (see the dark (blue) solid and dashed lines in Fig. \ref{fig21}(b)). This means that both stationary solutions of the time-evolution equation for $S(k;t)$, Eq. (\ref{relsigmadif2pp}), exist simultaneously in the interval $ T_s(\phi)\le T\le T_c(\phi)$. However, it is not difficult to see \cite{nescgle3} that under these circumstances, $S_a(k;\phi,T)$ is the only possible long-time asymptotic limit of $S(k;t)$. Thus, although $S^{(eq)}(k;\phi, T)$ exists and is well-defined in the interval $ T_s(\phi)\le T\le T_c(\phi)$, it does not have any influence in the structural relaxation of the system for quenches in that interval. This is why the spinodal divergence of $S^{(eq)}(k=0;\phi, T)$ bears no relevance on the long-time arrested structure of the system, and hence, the spinodal temperature $T_s$ ceases to have any dynamical significance in the  regime $\phi >\phi_b$. }

\textcolor{black}{ This means that if we revisit the predicted dynamic arrest diagram in Fig. \ref{fig3b}, we might as well  remove completely the dashed portion of the spinodal curve to indicate its dynamic irrelevance and to make clear the predicted continuity of the glass phase throughout the region of state space below the curve $T_c(\phi)$. In the high-packing/high-temperature regime this glass phase corresponds to the ``hard-sphere'' or ``repulsive'' high-packing glasses that result from compressing hard spheres to $\phi_g^{HS}\approx 0.58$, whereas in the low-packing/low-temperature regime it corresponds to ``attractive'' glasses, in which the intense (compared with $k_BT$) attractive interactions allow the formation of macroscopically homogeneous spanning clusters with large microscopic heterogeneities involving high local packing coexisting with voids of similar dimensions, to allow mean packing fractions much lower than $\phi_g^{HS}\approx 0.58$. This predicted scenario is fully consistent, at least qualitatively, with the existence of the glass region of the experimental scenario reported in Refs.  \cite{cardinaux,gibaud}, and referred to there as region III. }

\section{Predicted scenario and agreement with experiment.}\label{sectionIII.6}

\textcolor{black}{ In this section we shall actually revisit and update the dynamic arrest diagram of Fig. \ref{fig3b}. The resulting  updated diagram will constitute the core of the scenario predicted by the NE-SCGLE theory, whose connection  with the reported experimental scenario on the arrested spinodal decomposition of lysozyme solutions will also be discussed. Let us thus incorporate the two most relevant findings of the previous section, in the dynamic arrest diagram of Fig. \ref{fig3b}. The first is concerned with the dynamic irrelevance of the dashed portion of the spinodal curve. This leads to the predicted continuity of the glass phase throughout the region of state space below (and to the right of) the curve $T_c(\phi)$. }

\begin{figure}
\begin{center}
\includegraphics[scale=.28]{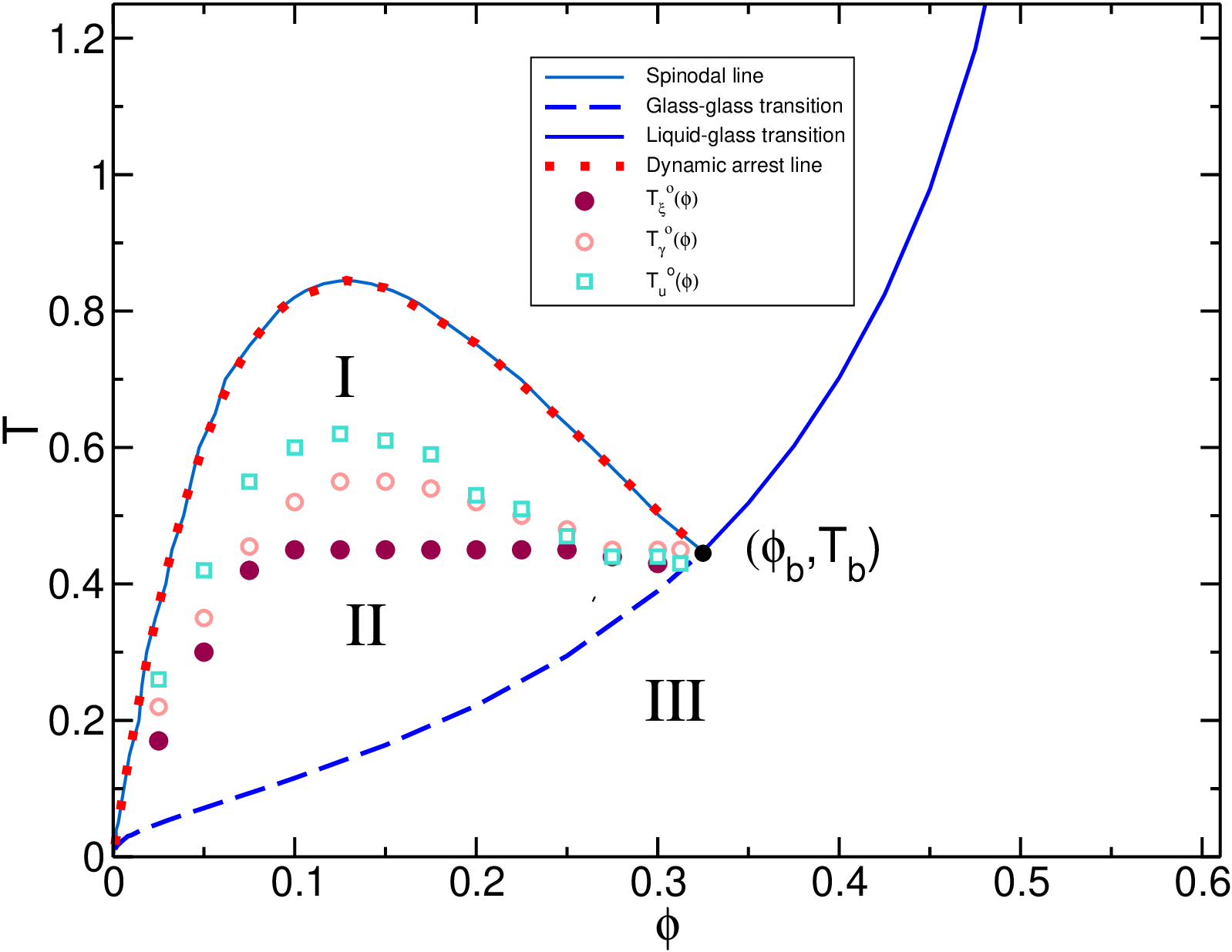}
\caption{\textcolor{black}{Dynamic arrest diagram with the crossover line $T_0(\phi)$ determined by the parametrization of the transition from exponential to power law of: the typical size $\xi_a(\phi, T)$ of the heterogeneities ($T^{0}_{\xi}$), the  square localization length $\gamma_a(\phi, T)$ ($T^{0}_{\gamma}$), and the  the material time $u_a(\phi,T)$ ($T^{0}_{u}$). Each of these three curves provide an approximate location of the crossover from full phase separation (region I) to gel formation (region II). Region III corresponds to the formation attractive glasses.}}
\label{fig90}
\end{center}
\end{figure}

\textcolor{black}{The second refers to the existence of a crossover temperature $T_0(\phi)$ at which several properties change their temperature dependence from an exponential to a power-law that  diverges at the spinodal line $T_s(\phi)$.  In Fig. \ref{fig90} we present the updated dynamic arrest diagram. There we have indicated the location of the crossover line  $T_0(\phi)$ obtained in the three manners described above (i.e., based on the $T$-dependence of  $\xi_a(\phi, T)$, of $\gamma^*_a(\phi, T)$, and of $u_a(\phi, T)$). We do not observe, but we did not expect, any quantitative agreement among these three results for $T_0(\phi)$. However, it is clear that within their quantitative differences, the three of them suggest a common qualitative and semiquantitative scenario. In fact, their spreading may serve to remind us that we are talking about a crossover between two limiting conditions, and not about a sharp transition between two well-defined states (like in the liquid-glass transition line). }

\textcolor{black}{The referred two limiting conditions are, for shallow quenches ($T\lesssim T_s$),  the ineffectiveness of the dynamic arrest conditions to prevent full gas-liquid separation, whereas for deeper quenches, slightly above the glass-glass transition ($T\gtrsim T_c(\phi)$), the effectiveness of the condition of dynamic arrest to interrupt the process of spinodal decomposition at an early stage of its development. The result of this interruption cannot be other than the formation of glassy states whose asymptotic structure is strongly determined by the amplified heterogeneities represented by the frozen small-$k$ peak of $S_a(k)$. These highly heterogeneous structures contrast with the much more homogeneous structure of the attractive glasses formed at deeper quenches, below the glass-glass transition ($T< T_c(\phi)$), in which the relevant spatial heterogeneities are determined by the main peak of  $S_a(k)$, and not by its insignificant small-$k$ peak.  Besides the difference in the scale and degree of the spatial structural heterogeneity of the two glass phases, we can highlight another relevant difference, this time in the localization length  $\sqrt{\gamma^*_a(\phi, T)}$.  From the results of the previous sections we can see  that $\sqrt{\gamma^*_a(\phi, T)}$ has the Lindemann-like value $\sqrt{\gamma^*_a(\phi, T)}\approx 0.15$ for the glass phase right below $T_c(\phi)$. In contrast, for the glass phase right above $T_c(\phi)$, the corresponding value of $\sqrt{\gamma^*_a(\phi, T)}$ is about ten times larger, thus indicating a sudden increase in the degree of heterogeneity upon crossing the glass-glass transition line $T_c(\phi)$. It is, of course, very difficult not to associate the fluffier glass phase above $T_c(\phi)$ with gel states, and the (rather loosely defined) crossover line $T_0(\phi)$ with the corresponding gel line.}

\textcolor{black}{In fact, if we now compare the topology of our updated theoretical diagram with its experimental counterpart in Fig. 4 of Ref. \cite{cardinaux} (or Fig. 1(a) of Ref. \cite{gibaud}), we immediately recognize that the qualitative mapping of one onto the other is impressive. To enhance this similarity with the experimental diagram, here we have also labeled, as in  Refs. \cite{cardinaux,gibaud}, the three kinetically distinct regions of the  theoretical diagram as I, II and III, since the interpretation of each of these theoretical regions, as discussed in this section, are conceptually consistent with the description in Refs. \cite{cardinaux,gibaud}  of the corresponding  regions experimentally determined in the lysozyme system. The most relevant feature shared by the predicted and the experimental scenarios, is the fact that the ``ordinary'' liquid-glass transition penetrates the unstable region as a novel glass-glass transition. For  $\phi >\phi_b$, this glass phase borders with the equilibrium liquid phase along the liquid-glass transition line $T_c(\phi)$, so that we expect that the predicted phenomenology of the corresponding liquid-glass transition will be essentially identical to that of the purely repulsive soft-sphere system studied in Ref. \cite{nescgle3}. Although there is still much to be understood about this ``ordinary'' glass transition, from the study in Ref. \cite{nescgle3} we have learned that as soon as we enrich the scenario provided by the dynamic arrest diagram (obtained solely from long-time asymptotic properties like $\gamma^*_a (\phi,T)$) with the kinetic dependence on waiting time, a number of important effects can be predicted, such as the aging of the structure and the dynamics, and the experimental impossibility of observing the sharp and discontinuous liquid-glass transition implied by the dynamic arrest diagram, which can only be observed as a blurred and continuous transition \cite{nescgle3}. }

\begin{figure}
\begin{center}
\includegraphics[scale=.28]{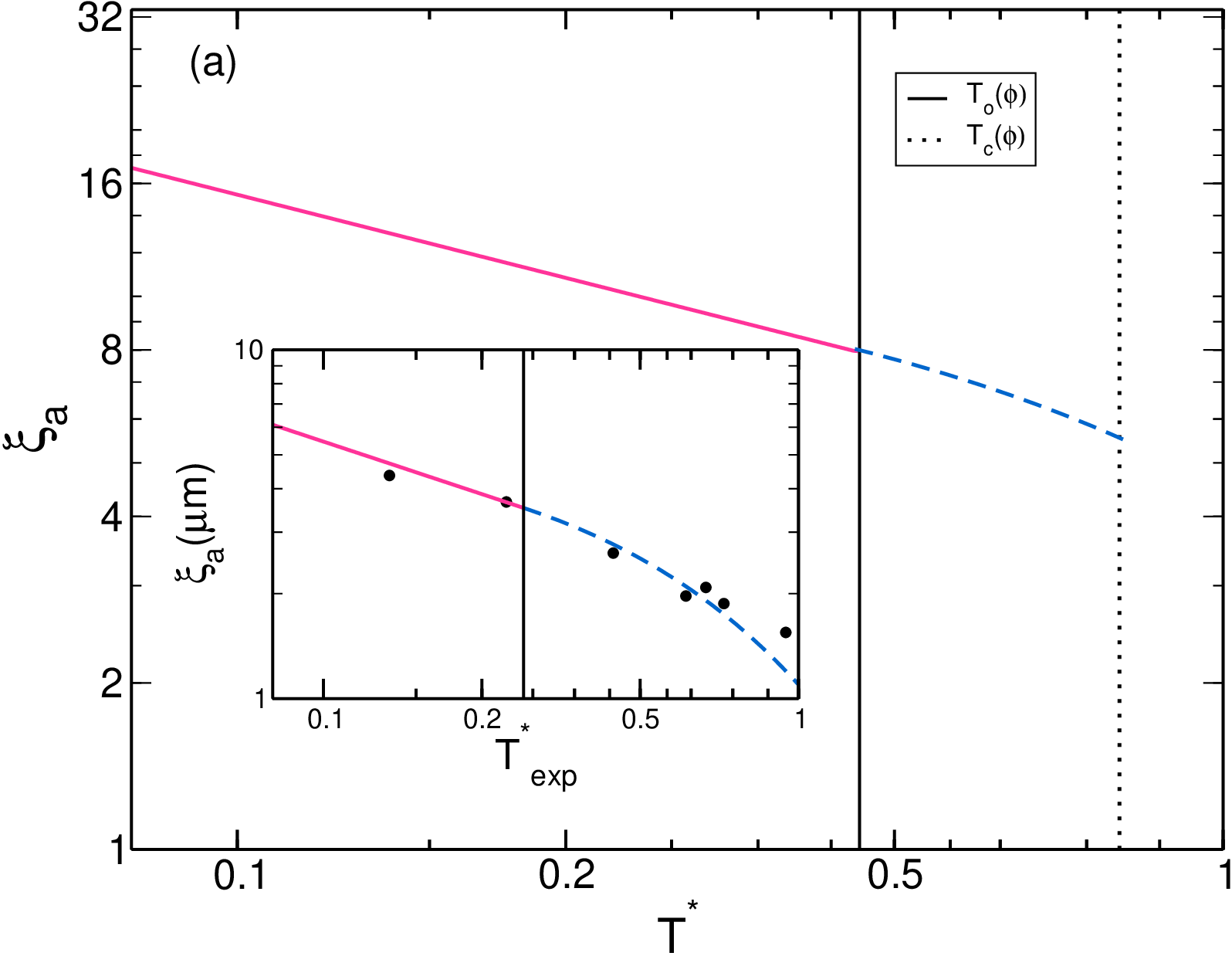}\hskip.5cm
\includegraphics[scale=.28]{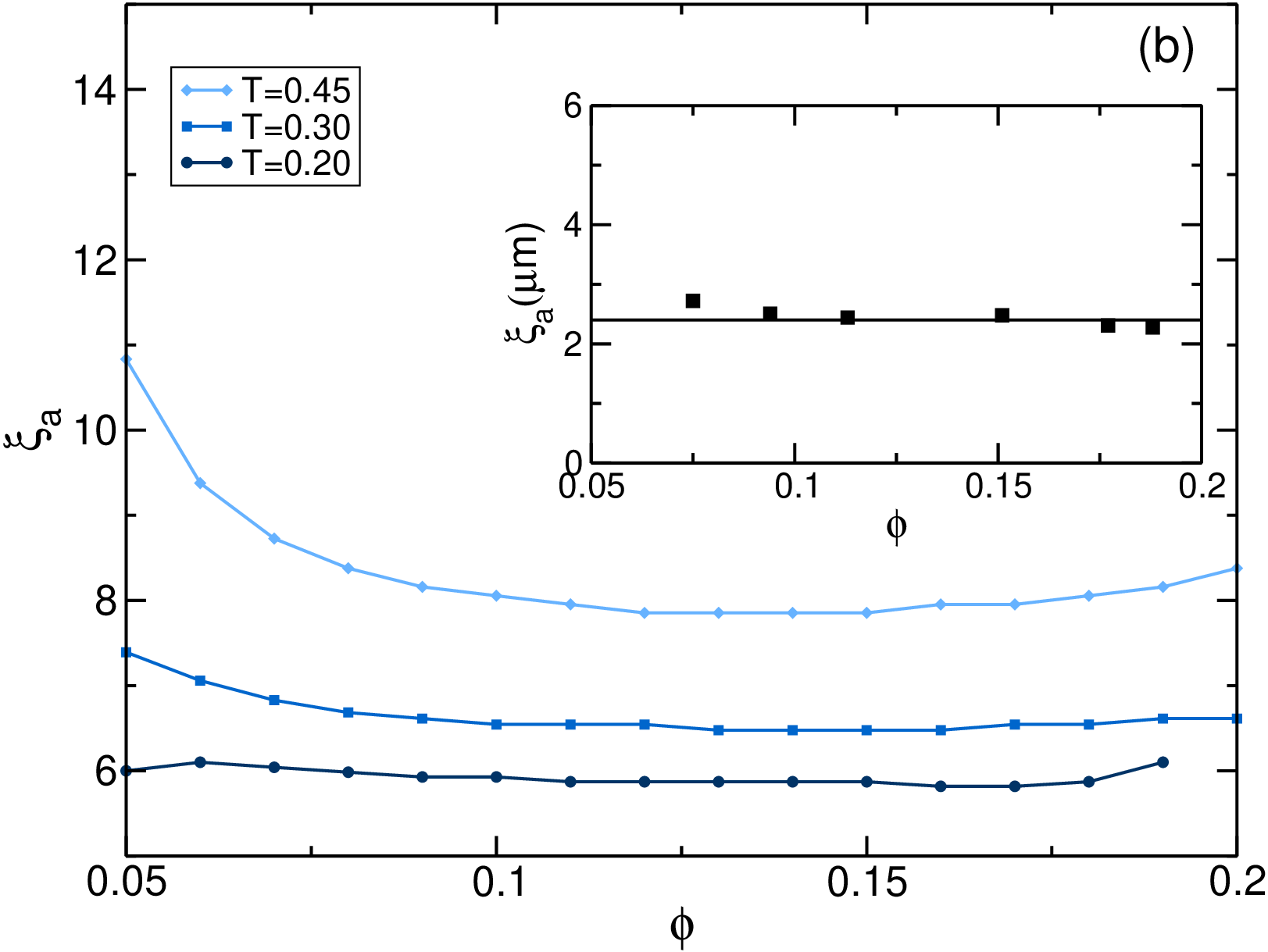}
\caption{\textcolor{black}{Comparison between the theoretical scenario and the experimental results shown in Figs. 3(b) and 4(c) of Ref. \cite{gibaud} for the characteristic length $\xi_a(\phi, T)=2\pi/k_{max}(\phi,T)$.  In the main panel of (a) we show the theoretical results for $\xi_a$ as a function of $T^*=(T_{s}-T)/T_{s}$ at $\phi=0.15$ for the HSAY model with $z=2$, parametrized according to Eq. (\ref{fitcsi}). In the inset the dots are the experimental data in Fig. 4(c) of Ref. \cite{gibaud} for the fast quench of the lysozyme solution at the same isochore, and  the composed solid-dashed curve is the fit of these data with the parametrization expression in Eq. (\ref{fitcsi}),  yielding $T_0 \ =12.6^{o}C$ (vertical solid line). Fig. (b) illustrates the experimental observation reported in Fig. 3(b) of Ref. \cite{gibaud} that for fixed final temperature $\xi_a(\phi, T)$ seems to be independent of volume fraction (inset). This feature is also qualitatively predicted by our theory, as illustrated by the results for  $\xi_a(\phi, T)$ along the isotherms $T=0.20, 0.30, 0.45$ plotted in the main panel (circles, squares and diamonds, respectively).  The experimental data in the insets were read from Figs. 3(b) and 4(c) of Ref. \cite{gibaud} with the permission of  IOP Publishing. } }
\label{fig97}
\end{center}
\end{figure}

\textcolor{black}{Let us finally mention other two additional points of comparison between our theoretical scenario and the experimental results of Schurtenberger and collaborators \cite{cardinaux,gibaud}. This group measured the temperature and density dependence of the  heterogeneities length scale $\xi_a(\phi, T)$ of the arrested structure of the  gels. They reported that for fixed volume fraction ($\phi=0.15$), $\xi_a(\phi, T)$ could be adjusted by $\xi_a(\phi, T)\propto [T_s-T]^{-\nu}$ with  $\nu=0.55 \pm 0.1$ \cite{gibaud}, and that $\xi_a(\phi, T)$, measured at fixed temperature, seemed to depend only weakly on volume fraction. Regarding the temperature dependence of  $\xi_a(\phi, T)$, in the main panel of Fig. \ref{fig97}(a),  we present the theoretical results for $\xi_a(\phi, T)$ in the HSAY model with $z=2$ along the isochore $\phi=0.15$ parametrized, as in Fig. \ref{fig9}, according to Eq. (\ref{fitcsi}), but plotted here as a function of the reduced temperature distance $T^*\equiv (T_s-T)/T_s$ to the spinodal curve. As in Fig. \ref{fig9}, the vertical (dotted and solid) lines represent, respectively, the glass-glass transition temperature $T_c(\phi)$ and the temperature  $T_0(\phi)$ indicating the crossover of the $T$-dependence of $\xi_a(\phi, T)$ from exponential (solid magenta portion) to diverging power-law (dashed blue portion). The prediction is that gels should be expected to form in the interval $T_c < T <T_0$ and to start to phase separate for temperatures above $T_0$. In the inset of Fig. \ref{fig97}(a)  we reproduce the log-log plot of the experimental data reported in Fig. 4(c) of Ref.  \cite{gibaud}  for the fast quench of the lysozyme solution at the isochore $\phi=0.15$. The composed (solid-dashed) curve is the parametrization expression in Eq. (\ref{fitcsi}), in which we fixed the exponent $\nu$ at its Ornstein-Zernike value $\nu=0.5$ and employed the reported experimental value of the parameters $T_s \ (=18^{o}C)$, $T_c\ (=-4.5^{o}C)$ and $\xi_a(\phi, T_c)\ (=1.05 \mu m)$, and varied $T_0$ to obtain the best fit of these experimental data, yielding $T_0 \ =12.6^{o}C$, a value 16 $\%$ below the experimental gel (or tie-line) temperature of 15$^{o}C$. }

\textcolor{black}{The  experimental observation that for fixed final temperature $\xi_a(\phi, T)$ seems to be independent of volume fraction is illustrated in the inset of Fig. \ref{fig97}(b), which plots the experimental data of Fig. 3(b) of Ref. \cite{gibaud}. In the main panel of our figure \ref{fig97}(b) we plot the theoretical $\xi_a(\phi, T)$ as a function of $\phi$ at fixed temperature, in the same  volume fraction window as the experimental report. We observe that indeed, in this particular window, the theoretical prediction  for $\xi_a(\phi, T)$ along the three isotherms presented can be said to depend rather weakly on  volume fraction. We must mention, however, that in a more extended window that includes lower densities, the spinodal divergence of  $\xi_a(\phi, T)$ will lead to a more complex scenario than the simple independence of  $\xi_a(\phi, T)$ on volume fraction, as already insinuated by the curve corresponding to $T=0.45$. Let us also stress that the comparison in Fig. \ref{fig97} between our theoretical predictions and the experimental results of Ref. \cite{gibaud} is only meant to illustrate the qualitative consistency between our theoretical predictions for a specific model system  and the results of an experimental study on the real system conceptually closest to our idealized model. Although the difference in the regime theoretically studied here (relatively long-ranged attractions, $z$=2) and the experimental system (much shorter-ranged attractions) prevents a more detailed direct comparison, it is clear that our theory is able to predict and describe the relevant measurable properties, and that the predictions are consistent with the available experimental reports. }

\section{Summary.}\label{sectionIV.3}

\textcolor{black}{In summary, in this paper we have introduced the Non-equilibrium Self-consistent  Generalized Langevin Equation theory of irreversible relaxation in liquids, as a general theoretical framework for the description of the non-equilibrium processes involved in the spinodal decomposition of suddenly and deeply quenched  simple fluids. The main result of the paper is the dynamic arrest diagram in Fig. \ref{fig90} of the previous section, which is the non-equilibrium analog of the equilibrium phase diagram in Fig. \ref{fig1}. This dynamic arrest diagram  complements the equilibrium scenario in Fig. \ref{fig1} with the kinetic perspective of the NE-SCGLE theory, which predicts the dynamic arrest  of a simple model liquid of attractive hard spheres suddenly and deeply quenched to its unstable region. Because of the non-linear and non-equilibrium nature of the phenomenon described, the initial interpretation of the specific predictions turned out to be rather subtle. However, the detailed analysis of several properties associated with the long-time asymptotic structure factor $S_a(k;\phi,T)$  provided a rather coherent picture, summarized by the resulting dynamic arrest diagram of  Fig. \ref{fig90}.}

\textcolor{black}{The first main feature of the predicted dynamic arrest diagram of attractive hard spheres is the continuity of Region III, corresponding to the glass phase below and to the right of the curve $T_c(\phi)$, which contains the high-packing/high-temperature ``repulsive'' glasses  and the low-packing/low-temperature ``attractive'' glasses.  For  $\phi >\phi_b$, this glass phase borders with the equilibrium liquid phase along the liquid-glass transition line $T_c(\phi)$. For $\phi <\phi_b$ this liquid-glass transition penetrates the unstable region of the gas-liquid coexistence as a glass-gel transition, with the gel phase constituting region II of the diagram. Finally, we associated the formation of gels with the early arrest of the growing spatial heterogeneities, whose size $\xi_a(\phi,T)=2\pi/k_{max}$ is determined by the position $k_{max}$ of the spinodal (small-$k$) peak of $S_a(k;\phi,T)$. The temperature dependence of this and other properties revealed the existence of two regimes, separated by a crossover temperature $T_0(\phi)$. This crossover line was conjectured to separate the region of gel formation from the region where the dynamic arrest is not able to prevent full phase separation.}

\textcolor{black}{This theoretical scenario emerged only form the analysis of long-time asymptotic structural properties, such as $S_a(k;\phi,T)$. The NE-SCGLE theory, however, provides abundant  additional information, including the detailed non-equilibrium evolution of the static and dynamic properties of the system at any instant after the quench. This kinetic and dynamic information leads to a richer and more detailed scenario whose description, however, is out of the scope of the present paper. Nevertheless, even without this important information, we related the available theoretical scenario with the results of the study of dynamically arrested spinodal decomposition in the experimental model system conceptually closest to the theoretical model analyzed here. As we demonstrated in the previous section, the most relevant predicted features are in satisfactory qualitative agreement with the experimental observations.}

\textcolor{black}{Of course, rather than closing the theoretical discussion of arrested spinodal decomposition, the results presented in this paper only open a number of question to further discussion. For example, we expect to expand this discussion to include the kinetic aspects not discussed here, as well as the dependence of the resulting dynamic arrest scenario on the main features of the interaction potential, such as its spatial range and its specific form (attractive Yukawa vs. square well, hard-sphere vs. soft sphere, Lennard-Jones, etc.). Similarly, we have not exhausted all the pertinent tests and comparisons of the theoretical predictions with available relevant simulations, such as the thorough simulation study of Testard, Kob and Berthier \cite{testardjcp}.} In future communications, however, we shall address these and other issues
that we could not properly treat in the present paper.

\vskip5cm

ACKNOWLEDGMENTS: This work was supported  by the Consejo Nacional
de Ciencia y Tecnolog\'{\i}a (CONACYT, M\'{e}xico), through grants
No. 182132 and 242364.

\vskip1cm

\appendix{\bf {Appendix. The NE-SCGLE theory: main approximations.}}\label{A}

Let us have in mind a simple
monocomponent liquid formed by $N$ Brownian particles in a volume
$V$, which diffuse with a self-diffusion coefficient $D^0$ between
collisions while interacting through a pair potential $u(r)$. The
NE-SCGLE theory was originally derived from a non-equilibrium
extension of Onsager's theory of thermal fluctuations
\cite{nescgle1}. This abstract theory provides the general structure of the time-evolution equations for the mean value and the covariance of the thermal fluctuations of the macroscopic variables of a non-equilibrium system. Applied to the local number concentration of particles  in our Brownian liquid, these two equations become
\begin{equation} \frac{\partial \overline{n}(\textbf{r},t)}{\partial
t} = D^0{\nabla} \cdot b(\textbf{r},t)\overline{n}(\textbf{r},t)
\nabla \beta\mu[{\bf r};\overline{n}(t)] \label{difeqdlpap}
\end{equation}
and
\begin{eqnarray}
\begin{split}
\frac{\partial \sigma(\textbf{r},\textbf{r}';t)}{\partial t} = &
D^0{\nabla} \cdot \overline{n}(\textbf{r},t) \
b(\textbf{r},t)\nabla \int d \textbf{r}_2
\mathcal{E}[\textbf{r},\textbf{r}_2;\overline{n}(t)]
\sigma(\textbf{r}_2,\textbf{r}';t) \\ & +  D^0{\nabla}' \cdot
\overline{n}(\textbf{r}',t) \ b(\textbf{r}',t)\nabla' \int d
\textbf{r}_2 \mathcal{E}[\textbf{r}',\textbf{r}_2;\overline{n}(t)]
\sigma(\textbf{r}_2,\textbf{r};t) \\ & -2D^0{\nabla} \cdot
\overline{n}(\textbf{r},t)  \ b(\textbf{r},t)\nabla
\delta(\textbf{r}-\textbf{r}'). \label{relsigmadif2pap}
\end{split}
\end{eqnarray}
which describe the non-equilibrium evolution of the mean value
$\overline{n}(\textbf{r},t)$ and the covariance
$\sigma(\textbf{r},\textbf{r}';t)\equiv \overline{\delta n
(\textbf{r},t)\delta n (\textbf{r}',t)}$ of the fluctuations
$\delta n(\textbf{r},t) = n(\textbf{r},t)-
\overline{n}(\textbf{r},t)$ of the local concentration profile
$n(\textbf{r},t)$ of a colloidal liquid.

There are two fundamental
elements entering in these equations. The first involves the
relevant thermodynamic information, and is the electrochemical
potential $\mu[{\bf r};n(t)]$ of one colloidal particle at
position {\bf r}, from which the thermodynamic stability function
$\mathcal{E}[{\bf r},{\bf r}';n]\equiv \left[ {\delta \beta\mu
[{\bf r};n]}/{\delta n({\bf r}')}\right]$
derives \cite{callen,keizer}  ($k_BT=1/\beta$ being the thermal
energy per particle). The other is the time-dependent local
mobility function $ b(\textbf{r},t)$, which in reality is also a
functional of $\overline{n}(\textbf{r},t)$ and
$\sigma(\textbf{r},\textbf{r}';t)$, thus introducing a non-linear
character to these equations. In what follows we provide a more
detailed definition of these objects.

\subsection{Thermodynamic elements.}\label{subsectionII.1}

The electrochemical potential $\beta\mu[{\bf r};n(t)]$ is  an
ordinary function of the particle's position {\bf r} and a
\emph{functional} of the local concentration profile
$n(\textbf{r},t)$. Although not explicitly stated, $\mu[{\bf
r};n(t)]$ is also a functional of the temperature profile
$T(\textbf{r},t)$. Here, however, we shall only have in mind
uniform temperature profiles, $T(\textbf{r},t)=T(t)$, which will
play the role of the external control parameter describing the
imposed thermal treatment of the system. The functional dependence
of $\beta\mu[{\bf r};n] $ on the arbitrary profile $n(\textbf{r})$
embodies  the chemical equation of state of the system, and is a
fundamental thermodynamic input of the present theory. The
functional derivative $\mathcal{E}[{\bf r},{\bf r}';n]\equiv
\left[ {\delta \beta\mu [{\bf r};n]}/{\delta n({\bf r}')}\right]$
is in general an ordinary function of both, {\bf r} and ${\bf
r}'$, and a \emph{functional} of $n(\textbf{r},t)$. Evaluated in
the absence of external fields and at an uniform profile
$n(\textbf{r}) = n_0$, this two-point thermodynamic property only
depends on the distance $|\textbf{r}-\textbf{r}'|$ and its
functional dependence on the profile $n(\textbf{r})$ becomes an
ordinary dependence of the scalar argument $n_0$, i.e.,
$\mathcal{E}[\textbf{r},\textbf{r}';\overline{n}]=
\mathcal{E}_h(|\textbf{r}-\textbf{r}'|;n_0)$. The so-called direct
correlation function $c(r;n_0)$ is defined in terms of this
thermodynamic function by the relationship
$\mathcal{E}_h(r;n_0)=\delta (r)/n_0-c(r;n_0)$.

In order to simplify the analysis of the equations above, at this
point we introduce the ``local-homogeneity'' approximation, in
which we approximate this two-point thermodynamic property by
$\mathcal{E}[\textbf{r},\textbf{r}';\overline{n}(t)]=
\mathcal{E}_h(|\textbf{r}-\textbf{r}'|;\overline{n}(\textbf{r},t))$,
i.e., by the thermodynamic matrix evaluated at an uniform
concentration profile, whose constant value is the local and
instantaneous concentration $\overline{n}(\textbf{r},t)$. Under
these conditions we expect that the covariance can also be
approximated as $\sigma(\textbf{r},\textbf{r}';t)\approx
\sigma(|\textbf{r}-\textbf{r}'|;\textbf{r},t)$, so that Eq.
(\ref{relsigmadif2pap}) can also be written as
\begin{eqnarray}
\begin{split}
\frac{\partial \sigma(k;\textbf{r},t)}{\partial t} = & -2k^2 D^0
b(\textbf{r},t)\overline{n}(\textbf{r},t)
\mathcal{E}_h(k;\overline{n}(\textbf{r},t)) \sigma(k;\textbf{r},t)
\\ & +2k^2 D^0 b(\textbf{r},t) \overline{n}(\textbf{r},t), \label{relsigmainhomoap}
\end{split}
\end{eqnarray}
where $\sigma(k;\textbf{r},t) \equiv \int d^3 k
e^{i\textbf{k}\cdot \textbf{x}} \sigma(|\textbf{x}|;\textbf{r},t)$
and $\mathcal{E}_h(k;\overline{n}(\textbf{r},t)) \equiv
(2\pi)^{-3}\int d^3 k e^{-i\textbf{k}\cdot \textbf{x}}
\mathcal{E}_h(|\textbf{x}|;\overline{n}(\textbf{r},t))$.

To actually solve these equations,  we must indicate the specific nature of the
system, i.e., the specific inter-particle interactions,
embodied in the pair interaction potential $u(r)$, and the
applied external force $\textbf{F}(\textbf{r})=-\nabla \psi({\bf
r})$  acting on a particle at position \textbf{r}, whose effects
are introduced through the chemical equation of state. This is
written in general as \cite{evans},
\begin{equation}\label{1}
\beta\mu [{\bf r};n]  \equiv  \beta\mu^{*}(\beta) + \ln n({\bf r})
+ \beta \psi({\bf r})-c[{\bf r};n].
\end{equation}
The first two terms on the right side are the ideal gas
contribution to the chemical potential, whereas the term $-c[{\bf
r};n]$ contains the deviations from ideal behavior due to
inter-particle interactions. Thus, the functional dependence of
$c[{\bf r};n]$ on the concentration profile $n({\bf r})$  must be specified. One possibility
is to propose a theoretical approximation for this dependence,
which in the language of density functional theory \cite{evans} is
actually equivalent to proposing an approximate free energy
functional. We quote, for example, the simplest such
approximation, referred to as the Debye-H\"uckel or random phase
approximation, in which $c[{\bf r};n]$ is written as
$c^{(RPA)}[{\bf r};n]=-\beta \int d^3\textbf{r}'
u(|\textbf{r}-\textbf{r}'|)n({\bf r}')$. An approximate $c[{\bf
r};n]$ leads to a corresponding approximate thermodynamic matrix
$\mathcal{E}[{\bf r},{\bf r}';n]$. For example,
$\mathcal{E}^{(RPA)}[{\bf r},{\bf r}';n]=
\delta(\textbf{r}-\textbf{r}')/n({\bf r})+\beta
u(|\textbf{r}-\textbf{r}'|)$. Other more elaborate approximations
are available, however, in the literature of the equilibrium
theory of inhomogeneous fluids \cite{evans}.

We
have assumed so far that the external fields are static and
represented by $\psi({\bf r})$. There is, however, no fundamental
reason why we have to restrict ourselves to these conditions. In
fact, the general equations of the NE-SCGLE theory above can be
used, within the range of validity of the underlying assumptions,
to describe the response of the system to prescribed
time-dependent external fields $\psi({\bf r},t)$ or to programmed
thermal or mechanical constraints described by a time-dependent
temperature $T(t)$ of the reservoir and/or an imposed
time-dependent total volume $V(t)$ of the system. Here we shall
assume, however, that such time-dependent fields and constraints
could have been used to drive the system to a prescribed initial
state with mean concentration profile
$\overline{n}^{0}(\textbf{r})$ and covariance
$\sigma^{0}(k;\textbf{r})$, but that afterward the external fields
are set constant, $\psi({\bf r},t)= \psi({\bf r})$.

\subsection{Absence of external fields and spatial heterogeneities.}\label{subsectionII.2}

The NE-SCGLE Eqs. (\ref{difeqdlpap}) and (\ref{relsigmainhomoap}) are supposed to predict how the system relaxes to its final equilibrium state whose mean profile and covariance are
$\overline{n}^{eq}(\textbf{r})$ and $\sigma^{eq}(k;\textbf{r})$.
Describing this response at the level of the mean local
concentration profile $\overline{n}(\textbf{r},t)$ is precisely
the aim of the recently-developed \emph{dynamic} density
functional theory (DDFT) \cite{tarazona1, archer}, whose central
equation is recovered from our theory by setting
$b(\textbf{r},t)=1$ in Eq. (\ref{difeqdlpap}). DDFT has been applied
to a variety of systems, including the description of the
irreversible sedimentation of real and simulated colloidal
suspensions \cite{royalvanblaaderen}. Similarly, the
time-evolution equation for the covariance, within the additional
small-wave-vector approximation $\mathcal{E}_h(k)\approx
\mathcal{E}_0 +\mathcal{E}_2 k^2 +\mathcal{E}_4k^4$, can be
recognized as the kinetic equation that governs the irreversible
evolution of the static structure factor of a suddenly quenched
liquid, developed in the description of the early stages of
spinodal decomposition (see, for example, Eq. (2.11) of Ref.
\cite{langer}, in which $\mathcal{E}_4=0$, or Eq. (23) of Ref.
\cite{goryachev}). In fact, the main motivation of the present
work is to contribute to the development of the theory of spinodal
decomposition starting from the general NE-SCGLE theory just
introduced.

The most essential aspect of the irreversible process of spinodal
decomposition is the development of strong spatial
heterogeneities. These originate in the thermodynamic spinodal
instability of thermal fluctuations of certain wavelengths, whose
amplification generates strong spatial heterogeneities, leading
eventually to the full separation of the system in two separate
regions. At least in its initial regime, the main features of this
irreversible phase separation process must be exhibited by the
time-evolution of the two observable variables involved in the
present theory, namely, the mean value
$\overline{n}(\textbf{r},t)$ and the covariance
$\sigma(k;\textbf{r},t)$, as recognized since the early work of
Cahn and Hilliard \cite{cahnhilliard} and Cook \cite{cook}. Our NE-SCGLE equations actually contain Cook's time-evolution equation for the non-stationary static structure factor $S(k;t)$, but goes beyond this classical theory by incorporating the possibility of dynamic arrest.

To see this, let us imagine that we subject our system to a prescribed protocol
of thermal and/or mechanical treatment, described by the chosen
temporal variation  $T(t)$ and $V(t)$ of the reservoir temperature
and total volume, but \emph{in the absence} of external fields, $
\psi({\bf r},t)=0$. We also assume that the thermal
diffusivity of the system is sufficiently large that any
temperature gradient is dissipated almost instantly (compared with
the relaxation of chemical potential gradients), so that the
temperature field inside the system remains uniform and instantly
equal to the imposed temperature of the reservoir, $T({\bf r},t)=T(t)$.

In contrast, since the essence of spinodal decomposition is the
development of strong spatial heterogeneities, in this case we
cannot in general assume that the mass diffusivity of the system
is sufficiently large that any gradient of chemical potential can
be dissipated almost instantly. If this were the case, the density
field $\overline{n}(\textbf{r},t)$ inside the system would always
remain uniform and instantly equal to the imposed bulk density,
i.e., $\overline{n}({\bf r},t)=\overline{n}(t)\equiv N/V(t)$.
Under such conditions, rather than solving Eqs. (\ref{difeqdlpap})
and (\ref{relsigmainhomoap}) for $\overline{n}({\bf r},t)$ and
$\sigma(k;\textbf{r},t)$,  we would only have to solve Eq.
(\ref{relsigmainhomoap}) for $\sigma(k;\textbf{r},t)$, with
$\overline{n}({\bf r},t)$ substituted by its imposed bulk value
$\overline{n}(t)$, which now becomes a control parameter. Although
the resulting scheme might seem too gross for the description of
the spontaneous spatial heterogeneities represented by the
neglected deviations $\Delta\overline{n}(\textbf{r},t) \equiv
\overline{n}(\textbf{r},t) -\overline{n}(t)$ (which are definitely
non-zero), here we shall see that it provides a welcome mathematical and numerical simplification of  Eq.
(\ref{relsigmainhomoap}) for $\sigma(k;\textbf{r},t)\approx \sigma(k;t)$, which may now be written as
\begin{equation}
\frac{\partial \sigma(k;t)}{\partial t} = -2k^2 D^0
\overline{n}(t) b(t) \mathcal{E}_h(k;\overline{n}(t)) \sigma(k;t)
+2k^2 D^0 \overline{n}(t)\ b(t). \label{relsigmahomogeneaap}
\end{equation}

\subsection{Determination of the $t$-dependent mobility $b(t)$.}\label{subsectionII.3}

As explained in Refs. \cite{nescgle1} and
\cite{nescgle2}, the time-dependent mobility $b(t)$ is in reality a \emph{functional} of $\sigma(k;t)$, and is determined by the generalized Einstein relation
\begin{equation}
b(t)= [1+\int_0^{\infty} d\tau\Delta{\zeta}^*(\tau; t)]^{-1},
\label{bdtap}
\end{equation}
with the $t$-evolving, $\tau$-dependent friction function
$\Delta{\zeta}^*(\tau; t)$ representing the effects of the interparticle interactions on the total friction coefficient of a representative tracer particle. The functional dependence of $\Delta{\zeta}^*(\tau; t)$ on $\sigma(k;t)$ introduces the essential non-linearities of the problem, and this functional dependence can only be determined in an approximate manner.

If we were dealing with an equilibrium system, we could simply adopt the corresponding approximation proposed by mode coupling theory or by the equilibrium version of the SCGLE theory. The non-equilibrium SCGLE theory extends the latter to non-equilibrium conditions, thus leading to an approximate expression for $\Delta{\zeta}^*(\tau; t)$, namely,
\begin{equation}
\begin{split}
  \Delta \zeta^* (\tau; t)= \frac{D_0}{24 \pi
^{3}\overline{n}(t)}
 \int d {\bf k}\ k^2 \left[\frac{ S(k;
t)-1}{S(k; t)}\right]^2  \\ \times F(k,\tau; t)F_S(k,\tau; t)
\end{split}
\label{dzdtquenchap}
\end{equation}
in terms of the non-stationary static structure factor $S(k;
t)\equiv \sigma(k; t)/\overline{n}(t)$ and of the collective and
self non-equilibrium intermediate scattering functions $F(k,\tau;
t)$ and $F_S(k,\tau; t)$.

In Ref. \cite{nescgle1} it was indicated how to extend the exact memory-function expressions for $F^{eq}(k,\tau)$ and $^{eq}F_S(k,\tau)$ to non-equilibrium conditions, thus leading to the corresponding exact memory-function expressions for $F(k,\tau;t)$ and $F_S(k,\tau; t)$. It is at the level of the corresponding memory functions where one has to introduce the essential approximations, also extending the arguments of the SCGLE theory to non-equilibrium conditions. This converts the exact memory function expressions for $F(k,\tau;t)$ and $F_S(k,\tau; t)$ into the following, approximate memory-function equations, written in terms of the Laplace transforms (LT) $F(k,\tau; t)$ and
$F_S(k,\tau; t)$, as
\begin{gather}\label{fluctquenchap}
 F(k,z; t) = \frac{S(k; t)}{z+\frac{k^2D^0 S^{-1}(k;
t)}{1+\lambda (k)\ \Delta \zeta^*(z; t)}},
\end{gather}
and
\begin{gather}\label{fluctsquenchap}
 F_S(k,z; t) = \frac{1}{z+\frac{k^2D^0 }{1+\lambda (k)\ \Delta
\zeta^*(z; t)}},
\end{gather}
with $\lambda (k)$ being a phenomenological ``interpolating
function" \cite{todos2}, given by
\begin{equation}
\lambda (k)=1/[1+( k/k_{c}) ^{2}], \label{lambdadkap}
\end{equation}
where the  phenomenological cut-off wave-vector $k_c$ depends on the system considered. In the particular case of an instantaneous isochoric quench, which is the subject of the present paper, these equations become Eqs. (\ref{relsigmadif2pp})-(\ref{lambdadk}) of Sect. \ref{sectionII}.


\vskip1cm


\begin{references}



\bibitem{callen} H. Callen, \emph{Thermodynamics}, John Wiley,
New York(1960).

\bibitem{mcquarrie}  D. A. McQuarrie {\em Statistical Mechanics},
Harper \& Row (New York, 1973).

\bibitem{keizer} J. Keizer, \emph{Statistical Thermodynamics of Nonequilibrium
Processes}, Springer-Verlag (1987).

\bibitem{casasvazquez0} G. Lebon, D. Jou, and J. Casas-V\'azquez,
\emph{Understanding Non-equilibrium Thermodynamics Foundations,
Applications, Frontiers}, Springer-Verlag Berlin Heidelberg
(2008).

\bibitem{cahnhilliard} J. W. Cahn and J. E. Hilliard, J. chem. Phys. \textbf{31}, 688
(1959).

\bibitem{cook} H. E. Cook, Acta Metall. \textbf{18}, 297 (1970).


\bibitem{furukawa} H. Furukawa, Adv. Phys.,  \textbf{34}, 703 (1985).

\bibitem{langer} J. S. Langer, M. Bar-on, and H. D. Miller,
Phys. Rev. A \textbf{11}, 1417 (1975).

\bibitem{dhont} J. K. G. Dhont, J. Chem. Phys. \textbf{105}, 5112 (1996).

\bibitem{goryachev} S. B. Goryachev, Phys. Rev. Lett. \textbf{72}, 1850 (1994).

\bibitem{luetalnature} P. J. Lu, E. Zaccarelli, F. Ciulla, A. B.
Schofield, F. Sciortino and D. Weitz, Nature \textbf{22}, 499
(2008).

\bibitem{sanz} E.  Sanz, M. E. Leunissen, A. Fortini, A. van
Blaaderen, and M. Dijkstra, J. Phys. Chem. B \textbf{112}, 10861
(2008).

\bibitem{cardinaux}  \textcolor{black}{F. Cardinaux, T. Gibaud, A. Stradner, and P. Schurtenberger, Phys. Rev. Lett. \textbf{99}, 118301 (2007).}

\bibitem{gibaud}   T. Gibaud and P. Schurtenberger, J. Phys.:
Condens. Matter \textbf{21}, 322201 (2009).

\bibitem{foffi} L. Di Michele, D. Fiocco, F. Varrato, S. Sastry, E. Eisera and G.
Foffi, Soft Matter \textbf{10}, 3633  (2014).

\bibitem{Gao}  \textcolor{black}{Y. Gao, J. Kim and M. E. Helgeson. Microdynamics and arrest of coarsening during spinodal decomposition in thermoreversible colloidal gels. Soft matter (2015) DOI: 10.1039/C5SM00851D.}

\bibitem{zaccarellireviewgels} E. Zaccarelli, J. Phys.: Condens.
Matter \textbf{19}, 323101 (2007).

\bibitem{heyeslodge} J. F. M. Lodge and D. M. Heyes,
J. Chem. Soc., Faraday Trans., \textbf{93}, 437 (1997).

\bibitem{testardjcp} V. Testard, L. Berthier, and W. Kob, J. Chem.
Phys. \textbf{140}, 164502 (2014).


\bibitem{foffi0}  \textcolor{black}{G. Foffi, C. De Michele, F. Sciortino, and P. Tartaglia, J. Chem. Phys. \textbf{122}, 224903 (2005).}


\bibitem{berthierreview} L. Berthier and G. Biroli, Rev. Mod. Phys.
\textbf{83} (2011).


\bibitem{goetze1}  W. G\"{o}tze, in {\em Liquids, Freezing and Glass Transition},
edited by J. P. Hansen, D. Levesque, and J. Zinn-Justin
(North-Holland, Amsterdam, 1991).

\bibitem{goetze2} W. G\"{o}tze and L. Sj\"ogren, Rep. Prog. Phys. {\bf
55}, 241 (1992).

\bibitem{goetze3}  W. G\"{o}tze and E. Leutheusser, Phys. Rev. A {\bf 11}, 2173 (1975).

\bibitem{goetze4} W. G\"{o}tze, E. Leutheusser and S. Yip, Phys. Rev. A {\bf
23}, 2634 (1981).

\bibitem{latz} A. Latz, J. Phys.: Condens. Matter, \textbf{12} (2000) 6353.

\bibitem{cugliandolo1}\textcolor{black}{ L. F. Cugliandolo and J. Kurchan, Phys. Rev. Lett. \textbf{71}, 173
(1993).}

\bibitem{rmf}  P.E. Ram\'{\i}rez-Gonz\'alez {\it et al.}, Rev. Mex. F\'{\i}sica
\textbf{53}, 327  (2007).

\bibitem{todos1} L. Yeomans-Reyna, M. A. Ch\'avez-Rojo,
P. E. Ram\'{\i}rez-Gonz\'alez, R. Ju\'arez-Maldonado, M.
Ch\'avez-P\'aez, and M. Medina-Noyola, Phys. Rev. E {\bf 76},
041504 (2007)

\bibitem{todos2} R. Ju\'arez-Maldonado {\it et al.}, Phys. Rev. E {\bf 76}, 062502 (2007).

\bibitem{nescgle0} P. E. Ram\'irez-Gonz\'alez and M. Medina-Noyola,
J. Phys.: Cond. Matter \textbf{21}: 504103 (2009).

\bibitem{nescgle1} P. E. Ram\'irez-Gonz\'alez and M. Medina-Noyola,
Phys. Rev. E \textbf{82}, 061503 (2010).

\bibitem{nescgle2} P. E. Ram\'irez-Gonz\'alez and M. Medina-Noyola,
Phys. Rev. E \textbf{82}, 061504 (2010).

\bibitem{nescgle3}  L. E. S\'anchez-D\'iaz, P. E. Ram\'irez-Gonz\'alez,
and M. Medina-Noyola, Phys. Rev. E \textbf{87}, 052306 (2013).

\bibitem{nescgle4}  L. E. S\'anchez-D\'iaz,
E. L\'azaro-L\'azaro, J. M. Olais-Govea and M. Medina-Noyola, J.
Chem Phys.  \textbf{140}, 234501 (2014).

\bibitem{oz1914}\textcolor{black}{ L. S. Ornstein and F. Zernike, Proc. K. Ned. Akad. Wet. \textbf{17}, 793
(1914).}

\bibitem{evans} R. Evans, Adv. Phys. {\bf 28}: 143(1979).

\bibitem{tarazona1} U. Marini Bettolo Marconi and P. Tarazona, J.
Chem. Phys. \textbf{110}, 8032 (1999); ibid., J. Phys.: Condens.
Matter \textbf{12}, A413 (2000)

\bibitem{archer} A. J. Archer and M. Rauscher, J. Phys. A \textbf{37}, 9325 (2004).

\bibitem{royalvanblaaderen} C. P. Royall et al.,
Phys. Rev. Lett. \textbf{98}, 188304 (2007).

\bibitem{gabriel} \textcolor{black}{L. L\'opez-Flores et al., EPL, \textbf{99}, 46001 (2012).}



\bibitem{sharmasharma} R. V. Sharma and K. C. Sharma, Physica A \textbf{89}, 213 (1977).

\bibitem{percusyevick}  J. K. Percus and G. J. Yevick,
Phys. Rev. {\bf 110}, 1 (1957).

\bibitem{verletweiss} L. Verlet and J. J. Weis {\em Phys. Rev. A} {\bf 5},
939 (1972).

\bibitem{hansenverlet} J. P. Hansen and L. Verlet, Phys. Rev.  {\bf  184} 151 (1969).



\end{references}
\end{document}